\documentstyle[12pt,draft,epsf]{article}

\newcommand \be  {\begin{equation}}
\newcommand \ee  {\end{equation}}
\newcommand \bea {\begin{eqnarray} \nonumber}
\newcommand \eea {\end{eqnarray}}

\newcommand \xiz {\xi^{(0)}}
\newcommand \xio {\xi^{(1)}}

\begin{document}

\title{On the $3d$ Ising Spin Glass}

\author{Enzo MARINARI$^{(a,b)}$, Giorgio PARISI$^{(c)}$,\\
{\small and}\\
Felix RITORT$^{(b,d)}$ \\[1.3em]
{\small $(a)$: NPAC and Department of Physics, }\\
{\small Syracuse University}\\
{\small Syracuse, NY 13244, USA}\\[0.3em]
{\small $(b)$: Dipartimento di Fisica and Infn,}\\
{\small Universit\`a di Roma {\it Tor Vergata},}\\
{\small Viale della Ricerca Scientifica, 00173 Roma, Italy}\\[0.3em]
{\small $(c)$: Dipartimento di Fisica and Infn,}\\
{\small Universit\`a di Roma {\it La Sapienza},}\\
{\small P. Aldo Moro 2, 00185 Roma, Italy}\\[0.3em]
{\small $(d)$: Departament de Fisica Fonamental,}\\
{\small Universitat de Barcelona,}\\
{\small Diagonal 645, 08028 Barcelona, Spain}\\
{\footnotesize \tt marinari@roma1.infn.it
        parisi@roma1.infn.it ritort@roma2.infn.it}\\[0.7em]}
\date{September 1993}
\maketitle

\vfill

\begin{flushright}
  {\bf SCCS 532}\\
  {\bf Roma Tor Vergata ROM2F/93/94}\\
  {\bf Roma La Sapienza 967-93}\\
  {\bf cond-mat/9310041 }\\
\end{flushright}

\newpage

\begin{abstract}
We study the $3d$ Ising spin glass with $\pm 1$ couplings. We
introduce a modified local action. We use finite size scaling
techniques and very large lattice simulations. We find that our data
are compatible both with a finite $T$ transition and with a $T=0$
singularity of an unusual type.
\end{abstract}

\vfill
\newpage

\section{Introduction}

Three dimensional spin glasses \cite{THE_BOOKS} are a fascinating
subject. Numerical simulations are here particularly interesting
\cite{OGIMOR,BHAYO1,OGIELS,BHAYO2,HETZEL,RIEGER}, since for such model
(the real thing) it is very difficult to obtain reliable analytical
results.  Up to date numerical simulations for the Ising case have
shown a phenomenology very similar to the experiments on real spin
glasses (for recent simulations and analytical results about, for
example, {\em aging} phenomena, see \cite{AGING}).  The study of small
size systems (up to a linear size $L=14$) has shown a reasonable
agreement with the predictions of broken replica theory, but it is
obscure how much information about the thermodynamic limit can be
inferred from the behavior of small systems. In particular one has to
be careful about extrapolating the pattern of replica symmetry
breaking from small to large lattices.  Here our aim has been to
reconsider the whole subject and try to clarify the emerging physical
picture at low temperature $T$.

We will deal with the problem of the nature and the existence of a
phase transition. A cursory look at the history of the subject is
useful. If we look at the period that begins when people investigated
first the subject of disordered spin systems we can easily establish
that there have been periodic oscillations, with periods of the order
of $7$ years.  Researchers in the field have been oscillating between
the credence that there is a sharp transition and the belief that
there are no transitions at all (as is maybe true in real glasses) and
that when lowering $T$ there is only a gradual freezing of the
dynamical degrees of freedom.  At the beginning theoreticians had (at
equal time) a different credence from the experimental researchers.
The two groups had a different frequency of oscillations, and now
there is consensus that the system undergoes some kind of phase
transition.

We have run long numerical simulations at various temperatures,
doing our best to distinguish among these two possibilities.  In
our analysis we have been very much inspired by the approach and
the doubts of Bhatt, Morgenstern, Ogielsky and Young
\cite{OGIMOR,BHAYO1,OGIELS,BHAYO2,HETZEL,RIEGER}, and over their results
we have tried to build and improve. We have found that the whole set
of our data is well compatible with the possibility that there is a
transition at a given non-zero $T$. Such a transition would be
characterized by a large value of the exponent $\gamma$, close to $2.5$
($\gamma$ is the usual susceptibility exponent, which will be defined
later in a more precise way). The whole set of data is also compatible
with a large set of possible reasonable functional dependencies, which
imply a transition temperature of zero.  Recent studies using improved
Monte Carlo techniques \cite{VALLEA,BERG,TEMPER} also find that doubts
about the existence of a finite $T$ critical behavior are justified
\cite{BERGSG}.  The difficulty of resolving among the two behaviors is
because a large value of $\gamma$ implies that the system is not far
from being at its lower critical dimension (at which, according to the
conventional wisdom, $\gamma \to \infty$).  The distinction among a
system at the critical dimension and a system very close to it is
particularly difficult to pin.  We believe however that we are not too
far from being able to resolve among the two models and that an
increase in the simulation time of one or two orders of magnitude can
clarify the situation. Of course precise theoretical predictions on the
behavior of spin glasses at the lower critical dimensions would be
invaluable.

We have been studying the $3d$ Ising spin glass, with $\pm 1$
couplings, but we have not used the standard first neighbor model.
Hoping for some gain we have simulated a slightly modified model with
second nearest and third nearest coupling. The reason for introducing
this model is that in the conventional model (on the usual cubic
lattice) the interesting pseudo-critical region is at very low
temperatures. In this region sensible numerical simulations are
extremely demanding in computer time, due to the extreme difficulty in
crossing even small barriers. We also believe that a systematic
comparison of results obtained with different Hamiltonians may be
useful in finding out those universal features that are independent
from the detailed form of the Hamiltonian.

In section (\ref{S_MODEL}) we define the model we use and the
quantities we measure.  In section (\ref{S_SMALL}) we present the
results obtained by using finite size scaling on small lattices
(from $4^3$ to $14^3$) while in section (\ref{S_LARGE}) we present the
results obtained on a large  lattice   $64^2 \times 128$.  Finally
in section (\ref{S_END}) we  present our conclusions.

\section{The Model and the Observable Physical Quantities
\protect\label{S_MODEL}}

We consider a three dimensional Ising spin glass model on a body
centered cubic lattice. In this model the lattice sites are
labeled by and integer valued three dimensional vector $i$. The
spins are defined on each lattice point and take the
values $-1$ or $1$.

The Hamiltonian of the model (with couplings $J_{i,k}$ that can
take the three values $0$ and $\pm 1$)
is

\be
  H[\sigma] \equiv
  -{1 \over 2} \sum_{i,k} J_{i,k} \sigma_i \sigma_k\ .
\ee

\noindent  The couplings $J$ may be zero or take randomly a value
$\pm 1$.  In the simplest version of the models $J_{i,k}$ is
different from zero if and only if

\be
  |i-k| \equiv \bigl ((i_x-k_x)^2 + (i_y-k_y)^2
  + (i_z-k_z)^2 \bigr )^{\frac{1}{2}} \ \le\  r\ \ .
\ee

\noindent
Different models may be obtained by changing the value of $r$. In
the limit $ r \to \infty $ we recover the infinite range SK model,
while for $r=1$ we define the usual short range nearest
neighbor model. In this paper we will discuss the model with
$r=3^{1/2}$, which corresponds to have $J \not= 0$ when all the
following three conditions are satisfied:

\be
  |i_x-k_x| \le 1,\ \ |i_y-k_y| \le 1,\ \ |i_z-k_z| \le 1\ ,
\ee

\noindent and $|i-k| \not= 0$. A crucial parameter in the model is the
effective coordination number $z$, which is the number of spins that
interact with a given spin (for $r=1$, $z=6$; for $r=3^{\frac{1}{2}}$,
$z=26$). For large values of $z$ the energy is proportional to
$z^{1/2}$. On a Bethe lattice (which is a refined mean field
approximation) the critical temperature may be computed exactly and
one finds that

\be
  (z-1)\tanh(\beta_{Bethe})^2=1\ .
\ee

In this approximation one finds $T^{(z=6)}_{Bethe}=2.08$ and
$T^{(z=26)}_{Bethe}=4.93$.  One difficulty with the
original $r=1$ model is that the hypothetical critical temperature
is small (about $1.1$). Since under a single spin-flip the
minimum change of the energy is $4$, such a low value of the
critical temperature implies a very small acceptance rate (about
$2\%$) for Monte Carlo steps in which we try to change the energy.
This effect should disappear for the $r=3^{1/2}$ theory.  Moreover
a different form of the lattice action may be useful to disentangle
the lattice artifacts from the universal behavior.

In the particular case of the $3d$ Ising spin glass a large value of
$z$ should increase the system reminiscence of to the infinite range
model. In a system at the lower critical dimension for high
values of $z$ we should see a sharp change of behavior from
the predictions of the mean field theory to the asymptotic low
energy behavior.

In order to define interesting observable quantities it is convenient
to consider two replicas of the same system ($\sigma$ and $\tau$). The
total Hamiltonian reads

\be
  H= H[\sigma]+H[\tau]\ .
\ee

\noindent  For the two replica system we can define the {\em
overlap}

\be
  q_i\equiv \sigma_i \tau_i\ ,
  \protect\label{E_OVER}
\ee

\noindent which will play a crucial role in our analysis. We will
introduce the correlation function of two $q$'s as follows:

\be
  G(i) \equiv \sum_k \langle q_{i+k} q_k \rangle =
  \sum_k \langle \sigma_{i+k} \sigma_k \rangle
         \langle \tau_{i+k}   \tau_k   \rangle =
  \sum_k \langle \sigma_{i+k} \sigma_k \rangle ^2\ .
\ee

\noindent We can use this correlation function to define a
correlation length. From high temperature diagram analysis (or from
the field theoretical approach) we expect that for large separation

\be
  G(i) \ \sim \  \frac{e^{-\frac{|i|}{\xi}}}{|i|} \ .
  \protect\label{E_G}
\ee

\noindent We can define an {\em effective mass} as

\be
   m(i) \equiv \log ( \frac{i\  G(i)}{(i+1)\ G(i+1)})\ .
\ee

\noindent We expect that at large $i$

\be
  \xi^{-1} = \lim_{i \to \infty} m(i)\ .
\ee

In our numerical simulations we have not measured the full $G(i)$.
We have measured the zero $2$-momentum Green functions

\be
  G^{(0)}(d) = \frac{1}{L_x \times L_y}\sum_{\rm x-y\  plane} G(i)\,
\ee

\noindent
where $d$ runs now only in one lattice direction. We will label with a
superscript $^{(0)}$ this kind of quantities. We have also measured
the site-site correlation function, but only summing over
contributions where one single coordinate change (by swapping the
lattice in a single chosen direction). Here the coordinate increment
has the form $(x,0,0)$. We will denote quantities defined in this way
with a superscript $^{(1)}$.

\noindent In a similar way in a finite volume we can introduce
the quantity

\be
  q \equiv {1 \over V} \sum_i q_i\ .
  \protect\label{E_Q}
\ee

\noindent In the infinite volume limit the {\em spin-glass
susceptibility} is defined as

\be
  \chi_o \equiv \overline{\lim_{V\to \infty}  V <q^2>}\ ,
\ee

\noindent where the upper bar denotes the average over the different
choices  of the disorder.

We expect the spin glass susceptibility and the correlation length
to diverge at the critical temperature with the critical
exponent $\gamma$ and $\nu$ respectively. Below the critical
temperature in the mean field approach $\chi_o$ is proportional to
the volume. More generally in the broken replica approach one finds
that

\be
  \lim_{V\to \infty}  V <q^m> =\int_0^1 dx\  q(x)^m\ ,
\ee

\noindent where $q(x)$ is the order parameter function defined in
ref. \cite{THE_BOOKS}.

In the high temperature phase no interesting physical predictions
can be obtained for the usual
magnetic susceptibility (divided by $\beta$) defined as

\be
  \chi \equiv \overline{\lim_{V\to \infty}  V <m^2>}\ ,
\ee

\noindent $m$ being the total instantaneous magnetization
($m \equiv {1 \over V}\sum_i \sigma_i$).  Gauge invariance implies
that at thermal equilibrium

\be
  \overline{\chi} =1\ .
\ee

At $T<T_c$ this equality is valid after summing over all
configurations with the correct Boltzmann weight. If we restrict the
sum only to configurations in a given equilibrium state this identity
does not apply.

\section{Finite Size Scaling\protect\label{S_SMALL}}

We will discuss here results obtained on small lattice sizes, in
situations where typically $L >> \xi$.  Since our goal is to
establish or disprove the existence of a critical behavior for
$T>0$ let us start by sketching the predictions of a finite-size
scaling analysis. If scaling is satisfied in the vicinity of a
critical point (at $T>0$), we expect

\be
  \chi_{o}\sim L^{2-\eta}\, f(\frac{L}{\xi})\ ,
  \protect\label{E_SCALING1}
\ee

\noindent where $\eta$ are the anomalous dimensions of the operator
$q$ defined in (\ref{E_Q}) and $\xi$ is the correlation length that
is expected to diverge at the critical temperature. Moreover, to establish
the existence of a finite critical temperature
it is useful to use the Binder parameter to locate the transition
point.  It is defined by

\be
  g(T) = \frac{1}{2}\ \bigl ( 3 -
  \frac{ \overline{\langle q^4 \rangle}   }
       {(\overline{\langle q^2 \rangle})^2}  )\ .
  \protect\label{E_BINDER}
\ee

If a finite $T$ phase transition exists we expect the curves $g(T)$
obtained for different lattice sizes to cross (asymptotically for
large enough lattices) at $T_c$. This is quite a
precise method to find the location of a critical point. For a $T=0$
singularity the same curves will merge in a
single curve as $T \to 0^+$. We will see that the possibility that
the exponent $\nu$ characterizing the divergence of the correlation
length is greater than $1$ makes arduous to distinguish between
these two cases.

As we have already discussed we want to distinguish among two
different scenarios. In one case there is a finite temperature
transition and the correlation length diverges like
$\xi\sim(T-T_c)^{-\nu}$.  In our finite size scaling analysis we
will use the large lattice best fit to $T_c$, $\gamma$ and $\nu$
from section (\ref{S_LARGE}). If a transition exists we have a
precise determination of the critical exponents and
parameters.

We should note here that if three is the lower critical dimension and
we have a $T=0$ singularity, it is not clear that the scaling relation
(\ref{E_SCALING1}) is satisfied. As we will discuss our results
suggest that if the scenario of a $T=0$ phase transition holds such
scaling behavior could not hold.  This violation of scaling appears in
the Heisenberg model in two dimensions and is a consequence of the
existence of the Goldstone modes. In the $O(N)$ symmetric Heisenberg
model for $N>2$ the correct scaling laws contains an effective
exponent:

\be
  \chi_{o}\sim L^{2-\eta(\frac{L}{\xi})}\, f(\frac{L}{\xi})\ ,
  \protect\label{E_AMIT}
\ee

\noindent where $\eta(0) = 0$. The dependence of the exponent
on $\frac{L}{\xi}$ is due to the instability of the $T=0$ fixed point.
In the $N=2$ case, there is no renormalization  of the coupling
constant (i.e. of the temperature). In the low temperature phase
one gets the simpler equation

\be
  \chi_{o}\sim L^{2-\eta(T)}\, f(\frac{L}{\xi})\ ,
\ee

\noindent where the function $\eta(T)$ is not an universal function.
Its value at the transition point, i.e., $\eta(T_c)$, is universal
and it is equal to ${1 \over 4}$.

We have simulated lattices with linear size $L=4,6,8,10,12$ from
$T=5.4$ down to the lowest temperature in which we were sure to have
thermalized ($T=2.6$ for $L=4$ and $T=3.6$ for $L=12$). We
have computed the overlap among two identical copies of the system,
defined in (\ref{E_Q}).

We have been careful in checking that we have really reached
thermal equilibrium. We have used as a basic criterion to check
that $\langle q \rangle$ was compatible with zero for each sample.

\begin{figure}
  \epsffile[120 206 565 440]{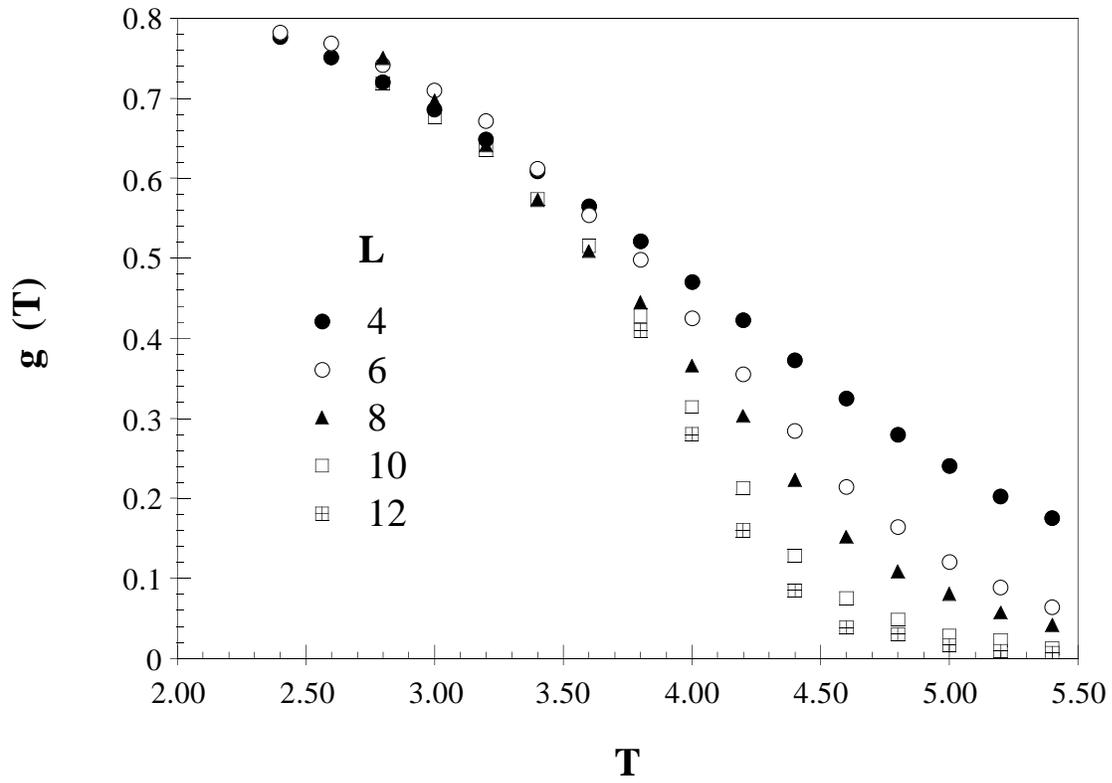}
  \caption[a]{\protect\label{F_FINITE1}
  The Binder parameter $g(T)$ as a function of the temperature
  $T$ for different lattice sizes.}
\end{figure}

We show in fig.~(\ref{F_FINITE1}) the Binder parameter defined in
eq.~(\ref{E_BINDER}), for different values of $T$. We cannot
distinguish any crossing, but we better see some merging of the
different curves.

\begin{figure}
   \epsffile[120 206 565 440]{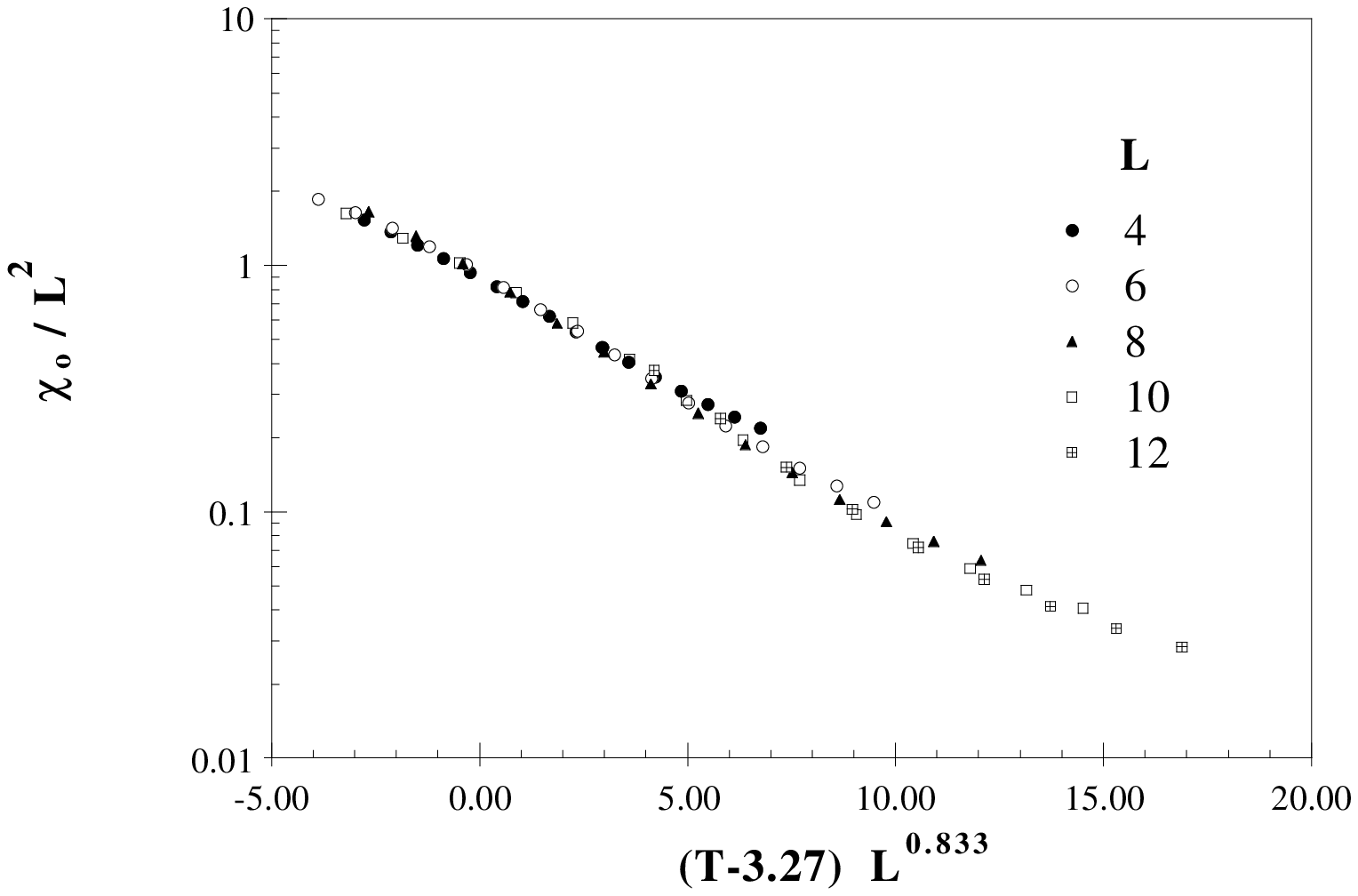}
  \caption[a]{\protect\label{F_FINITE2}
  The scaled overlap susceptibility $\frac{\chi_o}{L^2}$
  versus the scaled reduced critical
  temperature $(T-T_c)\,L^{\frac{1}{\nu}}$.}
\end{figure}

{}From the large lattice results (see section (\ref{S_LARGE})) we can
use the values  $T_c=3.27$ and $\gamma=2.4$ (see (\ref{E_CHIPARPOW}))
and $\eta \sim 0$ (see (\ref{E_CHIPARPOW})) to check the
consistency of the finite size behavior with a finite $T$
transition. In fig.~(\ref{F_FINITE2}) we plot $\frac{\chi_o}{L^2}$
versus $(T-T_c)\,L^{\frac{1}{\nu}}$. The data collapse on a
single curve, showing a good scaling behavior, on both sides of
$T_c$. It is already clear from these first data (illustrated in
figures (\ref{F_FINITE1}) and (\ref{F_FINITE2})) that it will be
exceedingly difficult to distinguish between the two candidate
critical (with $T_c = 0$ or $T_c \not= 0$) behaviors.

To understand  better what is happening in the pseudo-critical
region, for $T$ close to $3.3$, it is interesting to apply to the
model a magnetic field $h$. We expect $q$ to scale as
$h^{\frac{2}{\delta}}$. $\delta$ is related to $\eta$ by the
hyper-scaling relation

\be
  \delta=\frac{d+2-\eta}{d-2+\eta}\ .
  \protect\label{E_SCALING2}
\ee

\noindent In presence of $h$ the correct definition of the overlap
susceptibility requires subtraction of the connected part, i.e.,

\be
  \chi_o \equiv \lim_{V\to\infty}V(
  \overline{\langle q^2\rangle}-
  \overline{\langle q  \rangle}^2)\ .
  \protect\label{E_QH}
\ee

\noindent For a finite $T$ phase transition the scaling
relation (\ref{E_SCALING1}) is still satisfied, but now (we are
sitting at $T_c$) $\xi$ diverges like

\be
  \xi(h) \sim h^{-\frac{2(\delta +1)}{d\delta}}\ .
  \protect\label{E_SCALING3}
\ee

\noindent (\ref{E_SCALING3}) only depends on the critical exponent
$\eta$. Once we have measured $T_c$, and established that a finite
$T$ phase transition exists, we can use (\ref{E_SCALING3}) to
find $\eta$.

It turns out that the correct overlap susceptibility we have just
defined in (\ref{E_QH}) is not a good observable for checking scaling.
It depends on the first moment of $\langle q \rangle$ that is
affected by strong finite size corrections. This is because the
region of negative overlaps with $q<0$ is only suppressed in the
infinite size limit.  We have found preferable to study the
behavior of the non-subtracted $\overline{\langle q^2 \rangle}$,
i.e., of the overlap susceptibility defined in absence of $h$
divided times the volume. Here we expect the scaling
(\ref{E_SCALING1}) divided times $L^3$, i.e., a scaling with $L$
with the power $-(1+\eta)$.

\begin{figure}
  \epsffile[120 206 565 440]{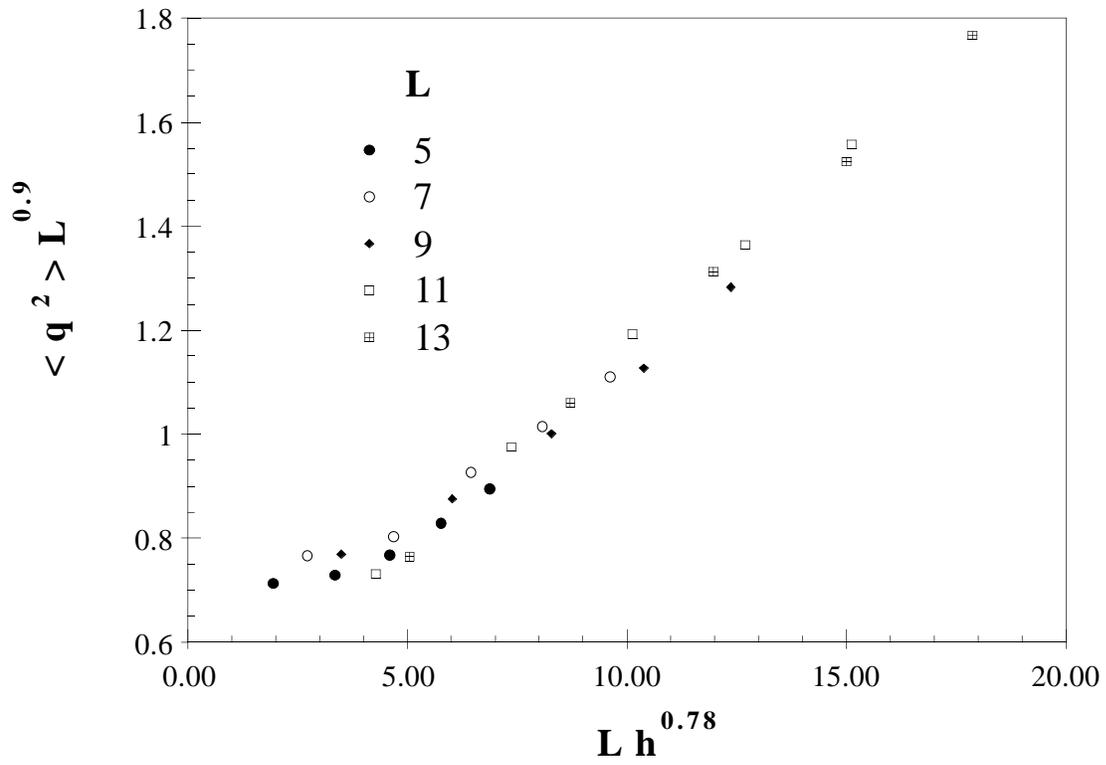}
  \caption[a]{\protect\label{F_FINITE3}
  The rescaled $\langle q^2 \rangle$ as a function of the
  rescaled applied magnetic field, for different lattice sizes.}
\end{figure}

We have run numerical simulations in presence of a magnetic field.
In figure (\ref{F_FINITE3}) we show  $\overline{\langle q^2 \rangle}$
for several lattice sizes $L=5,7,9,11,13$ and different values of
the magnetic field (ranging from $h=0$ up to $h=1.5$). Again
we find consistency with $T_c=3.27$. The preferred value for $\eta$
is negative and close to $-0.1$. Let us stress that all the finite
size scaling fits are not giving very precise predictions. There
are many free parameters, and that makes the fitting procedure
questionable. Still we should note that all the exponents we
find, when assuming a finite $T$ transition, are fully compatible
with the ones found for the $r=1$ model in the previous work of
references \cite{OGIMOR,BHAYO1,OGIELS,BHAYO2}.

\begin{figure}
  \epsffile[120 206 565 440]{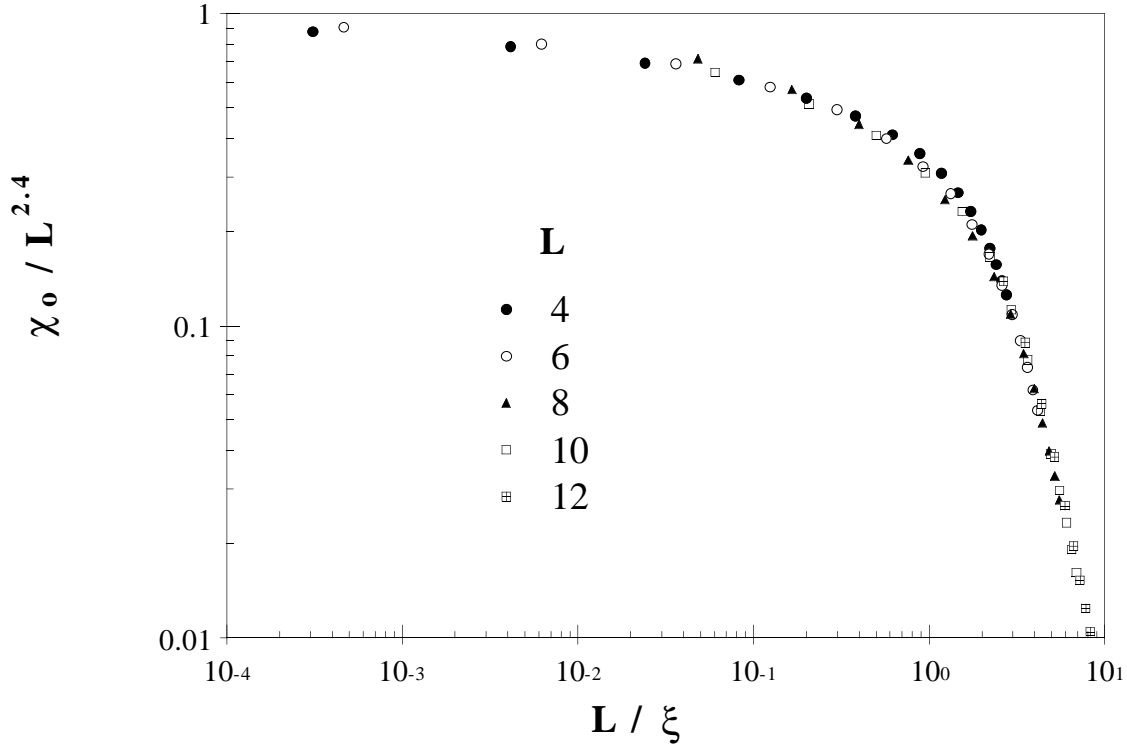}
  \caption[a]{\protect\label{F_FINITE4}
  The rescaled $\chi_o$ as a function of $\frac{L}{\xi}$. This
  scaling is compatible with a singularity at $T=0$.}
\end{figure}

As we already hinted the finite size scaling results are also
compatible with a $T=0$ singularity. We will use the best
value (\ref{E_CSIPAREXP}) of the parameters defined in
(\ref{E_CSIEXP}). In figure (\ref{F_FINITE4}) we show the rescaled
susceptibility $\chi_o$ (again without magnetic field, now) for the
different lattice sizes. The curves for different lattice
sizes scale tremendously well, and the comparison with figure
(\ref{F_FINITE2}) is instructive. This is, as we will discuss in
the following, fully compatible with the results obtained for the
large lattice size, in a regime where $\xi >> L$.

If the transition is at $T=0$ the usual scaling laws imply that the
correlation function at large distance behaves as $x^{-\zeta}$, with
$\zeta=d-2+\eta$.  When the ground state is not degenerate the $T=0$
correlation function goes to a constant value at large distance,
implying $\zeta=0$ and in $3d$ $\eta=-1$.  The value we estimate for
$\eta$ turns out to be not so close to $-1$, and using $\eta=-1$ does
not make our curves to scale.

Here we see two options. One possibility
is that $\zeta \not= 0$ in $3d$ Ising spin glasses (our best fit is
close to $\zeta\sim.6$). This possibility cannot be excluded. For
example in
$2d$ \cite{BHAYO2} $\zeta$ is estimated to be in the range $.2 \sim
.3$.  In our case, where the coupling constants $J$ take the values
$\pm 1$, the ground state is highly degenerate, and there are no
general a priori reasons for $\zeta=0$ to hold (however it has been
suggested in \cite{DK} that at the lower critical dimension we expect
indeed $\zeta=0$).  The other possibility is that to get good scaling
for $\eta$ we have to go at lower values of $T$. Here we have been
obliged to seat at not too low $T$'s and it is quite possible that
the value of $eta$ in this temperature range is quite different
from its zero temperature limit.

\begin{figure}
  \epsffile[120 206 565 440]{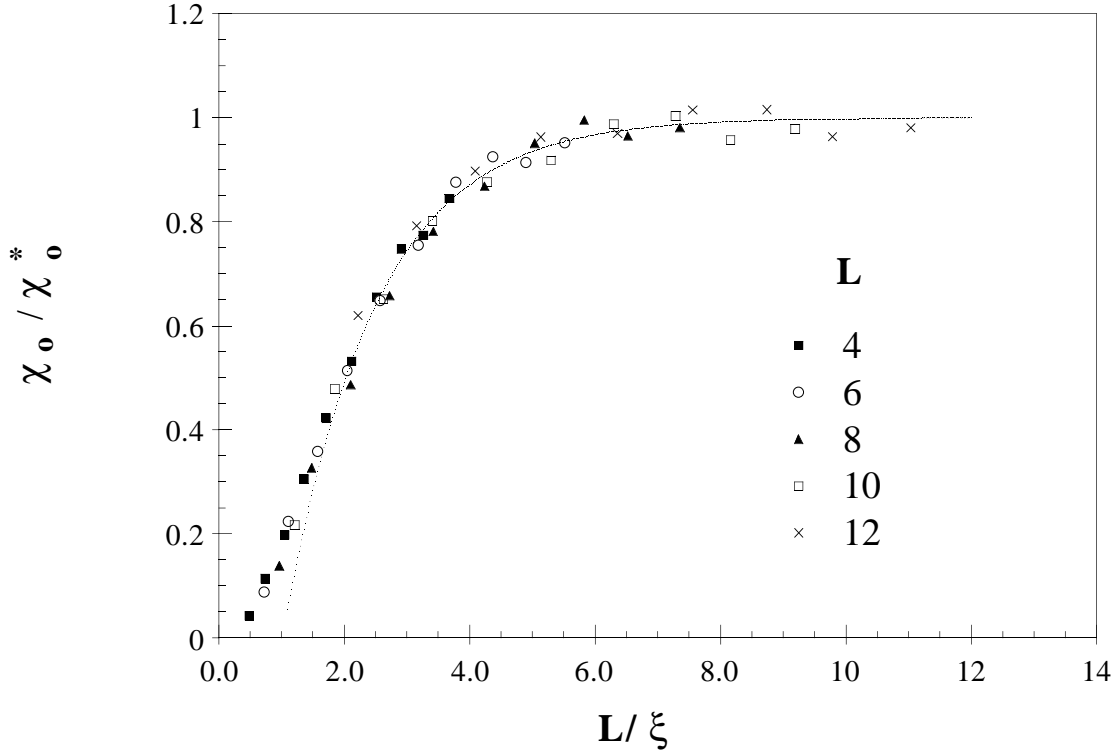}
  \caption[a]{\protect\label{F_FINITE5}
  The overlap susceptibility divided times the asymptotic large
  lattice value (which we denote here by $\chi_o^*$) versus
  $\frac{L}{\xi}$. The line is the best fit to the
  form (\ref{E_FINIFIT}). }
\end{figure}

In figure  (\ref{F_FINITE5}) we have tried to show the scaling behavior
in a suggestive form.  We plot $\frac{\chi_o(L)}{\chi_o(\infty)}$ as a
function of $L/\xi$ for the different lattice sizes.  The values of
$\chi_o(\infty)$ and $\xi$ are those discussed in the next
sections and computed on very large lattices (which we judge to be
free from systematic errors in our statistical precision). The data
smoothly collapse on a single curve.

{}From these data it is not clear if the $3d$ Ising spin glass undergoes
a finite $T$ phase transition (and mainly the puzzling behavior of
Binder cumulant seems to point toward something different). If we
assume a finite $T_c$ our predictions for the critical exponents agree
with those reported in the literature (for the first neighbor cubic
lattice model).

Though high temperature expansions predict a finite temperature
transition (which agrees with that found in numerical
simulations) we consider the compatibility of our data with a $T=0$
phase transition serious (and we will discuss this kind of
evidence in more detail in next section, when discussing our large
lattice results).

As we have already remarked the behavior of the Binder cumulant
below $T_c$ is different from what happens in normal spin
systems. It is also very different from what one measures in spin
glasses in high dimensions, and a few more comments are maybe in
order. Let us consider what happens in the usual  ferromagnetic
Ising case, by defining the function

\be
  g(T) \equiv \lim_{L \to \infty} g(T,L)\ ,
\ee

\noindent where here $g(T,L)$ is defined in terms of the moments of the
order parameter $m$, the total magnetization of the system.
In this non-disordered case we have that

\be
  {\rm for}\  T < T_c\ g(T) = 1\ ,\  {\rm for}\
  T >  T_c \ g(T) = 0\ .
\ee

\noindent Moreover the quantity $g_c \equiv g(T_c)$ is a function
of the dimensionality of the system. It increases when the dimension
decreases, and goes to $1$ at the lower critical dimension.

The situation is different in spin glass models. In this case in the
mean field approximation $g(T)$ is not trivial at low temperature. One
finds that below $T_c$

\be
  g(T) = \frac{3}{2} - \frac{1}{2}
  \frac{\int dx \ q(x)^4}
  {\bigl (\int dx \ q(x)^2\bigr )^2}
  = \frac{3}{2} - \frac{1}{2}
  \frac{\int dP(q)\ q^4}{\bigl ( \int dP(q)\ q^2 \bigr )^2}\ .
\ee

Using the mean field  expression  for the dependence
of $q(x)$ over $T$ one finds that

\be
  g_- \equiv \lim_{T \to T_c^-} g(T) = 1\ ,
\ee

\noindent but the function $g(T)$ is non trivial.
The statement $g \not= 1$ coincides with the fact that the $P(q)$ is
not equal to a $\delta$-function, and implies replica symmetry
breaking.  In the mean field approximation  no closed formula exists for
$g$, however one finds that qualitatively $g$ behaves as

\be
  1 - A \frac{T}{T_c}  (1-\frac{T}{T_c})\ .
\ee

\noindent In other words $g(T)$ vanishes linearly both at zero temperature
and at the critical temperature. For $T > T_c$ one still finds that $g(T)=0$.
Below the upper critical dimension ($d^{(u)}_c=6$) according to the
prediction of ref. \cite{DK} $g_-$ becomes different from $1$.
Slightly below $d=6$ the function $g(T)$ is not monotonous, but it is
possible that it becomes monotonous at sufficient low dimensions,
i.e., near three dimensions.  It is tempting to conjecture that near
the critical dimension one finds that $g_c$ becomes close to $g_-$. It
is difficult to assess quantitatively the values of these two
quantities. If we use our best estimate for $T_c$ we find $g_- \simeq
(.65 \pm .05)$, and a very similar value for $g_c$. We can only
tentatively conclude that:

\begin{itemize}

\item

The $L$ independence of $g(L,T)$ in the (pseudo)-low temperature phase
and the fact that $g(L,T)$ is different from $1$ is a clear signal
that replica symmetry is effectively broken in this region. Obviously
if there is no finite $T$ phase transition this symmetry breaking will
eventually disappear for very large lattices, but it will correctly
describe the physics of the system for large lattices with $L$ smaller
than the exponentially large correlation length $\xi$.

\item

The shape of the function $g(T)$ is in qualitative agreement with
the predictions of the renormalization group and it suggests that
the lower critical dimension is close to $3$ (and very probably
exactly $3$).

\end{itemize}

Let us now discuss in some detail the form of finite size effects.
This is very interesting, mainly since we have to plan larger
scale numerical simulations, and we want to be sure to optimize the
use of our computer time. We will describe here the strategy that
should eventually lead us to a numerical simulation in which we can
establish in a clear way which kind of singularity the $3d$ Ising
spin glass undergoes. For lattice sizes much larger than the
correlation length one finds that (in presence of periodic
boundary conditions) the finite volume corrections are
exponentially small. The leading correction can be computed in
perturbation theory, giving:

\be
  \chi_o(L)= \chi_o(\infty)
  \bigl (1-{\cal C}\  \xi^3 \lambda^2 e^{\frac{-L}{\xi}} \bigr)\ ,
\ee

\noindent where $\cal C$ is some computable constant,
and $\lambda$ is the coupling
constant of a $\phi^3$-like interaction in a field
theoretical framework. Close to the critical point  the usual
scaling laws imply that quantity $\xi^3 \lambda^2$ goes to a
constant. So we obtain:

\be
  \chi_o(L)= \chi_o(\infty)
  \bigl (1- {\cal C}\  e^{\frac{-L}{\xi}}
  +O(e^{\frac{-2L}{\xi}}) \bigr )\ .
\ee

\noindent We have fitted our data for the correlation length on
small lattices, divided over the large lattice result, as

\be
  \frac{\chi_o(L)}{\chi_o(\infty)}  \simeq (1- {\cal C} e^{\frac{-L}{\xi}})\ .
  \protect\label{E_FINIFIT}
\ee

The best fit works very well. We show it in figure
(\ref{F_FINITE5}). For a finite temperature transition
$\cal C$ is important. It is universal and in principle it can be
computed in a field theoretical renormalization approach.

These data are relevant since they are crucial for planning
simulations free of finite size effects on large lattices.
We see that if we require finite size effects to be smaller than
$1\%$ we need to have $L/\xi > 6$, while to reach a $10\%$
accuracy we can accept $L/\xi >3.5$.

In a similar way it is interesting to compute

\be
  K={\overline{\chi_o^2}-
      \overline{\chi_o}^2 \over \overline{\chi_o}^2}\ .
\ee

\noindent The quantity $K$ measures the  susceptibility system to
system fluctuations. We expect it to have similar properties to the
Binder cumulant $g$. In particular at low temperatures mean field
predicts that

\be
   K =
  \frac{1}{3} \Bigl(
  {\int dx \ q(x)^4 \over (\int dx \ q(x)^2)^2}-1 \Bigr) \ .
\ee

\noindent In other words mean field theory predicts that

\be
  \overline{\langle q^4\rangle - \langle q^2\rangle^2}=
  2 \bigl ( \overline{\langle q^2\rangle^2}
  -\overline{\langle q^2\rangle}^2 \bigr )\ .
\ee

The size dependence of $K$ can be used to estimate the number of
different realizations of the quenched disorder we need to
extract an accurate value of $\chi_o$.

\begin{figure}
  \epsffile[120 206 565 440]{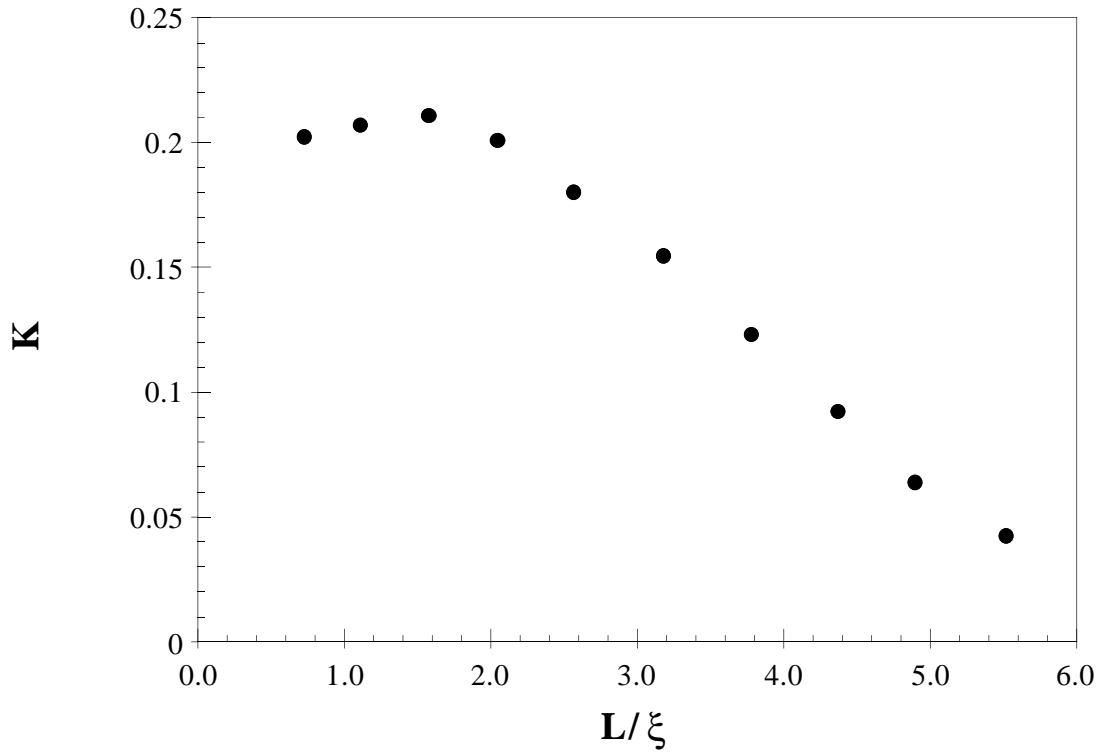}
  \caption[a]{\protect\label{F_FINITE6}
  $K$ as a function of $\frac{L}{\xi}$.}
\end{figure}

The measurament of $K$ is rather delicate because for each
system
we must know the value of $\langle q^2\rangle$ with high
accuracy.
In figure (\ref{F_FINITE6}) we plot $K$ as a function of $\frac{L}{\xi}$
for $L=6$.

\section{Large Lattice Results and Discussion\protect\label{S_LARGE}}

Our {\em large} lattice runs have been done on a $64\times 64\times
128$ lattice, on the $8192$ processor DECmpp at Syracuse NPAC.  We
have always studied the evolution of two replica of the system in
the same realization of the quenched disorder. In this way we have
been able to compute the overlap between two replica.

\begin{figure}
  \epsffile[22 206 565 690]{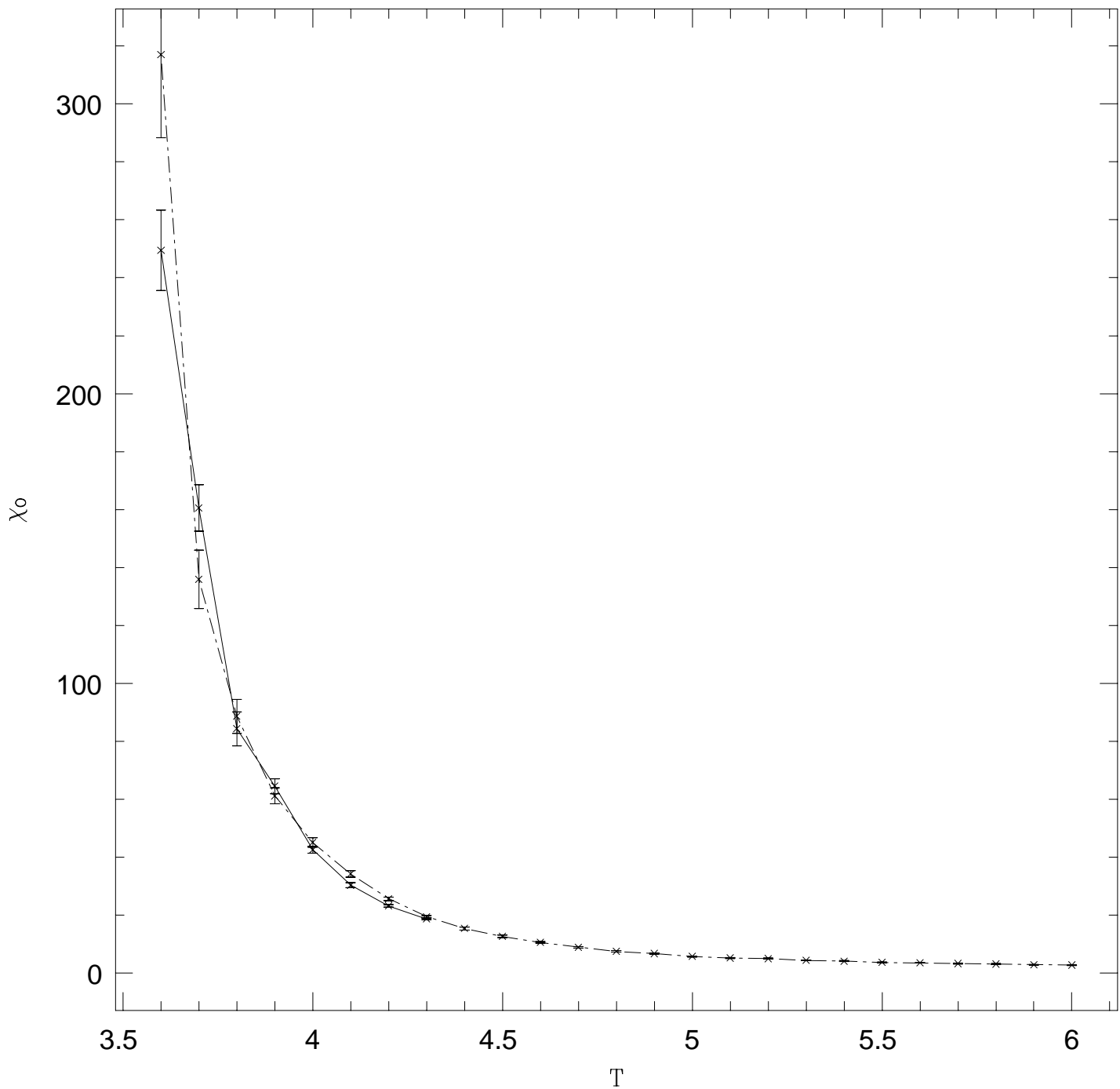}
  \caption[a]{\protect\label{F_2CHI}
  $\chi_o$, the overlap susceptibility, for the two realizations of the
  random couplings.}
\end{figure}

We have studied the behavior of the system for two different
realizations of the quenched random couplings.  We give in table
(\ref{T_SWE}) the details about the two series of runs (the number of
millions of sweeps is for each of the two replica we studied in a given
coupling
realization).

\begin{table}
  \begin{tabular}{|c||r|r|} \hline
    $T$ & Sample $1$ & Sample $2$ \\ \hline \hline
    $6.0 \to 4.4$ & $0.005 + 0.5$ &  \\ \hline
    $4.3 \to 3.8$ & $0.5 + 2.0$ &  $0.5 + 2.0$\\ \hline
    $3.7$ & $0.5 + 4.5$ &  $0.5 + 14.5$\\ \hline
    $3.6$ & $2.5 + 9.0$ &  $2.5 + 30.0$\\ \hline
  \end{tabular}
  \caption[a]{For the two different
  realizations of the couplings, the
  number of millions of MC sweeps we used. We give the number of thermalization
  sweeps, plus ($+$) the number of sweeps used for
  measuring.\protect\label{T_SWE}}
\end{table}

We studied two different realizations of the random noise mainly to
check the size of the fluctuations of $\chi_o$. We wanted to be sure
that even for our $T$ point closer to criticality ($T=3.6$) sample
to sample fluctuations are not too dramatic. In fig.~(\ref{F_2CHI})
we  show that in the worst case the two results for $\chi_o$
deviate of less then two standard deviation (in this and in the
following figures the smooth lines just join the Monte Carlo data
points with straight segments).  But we know from our binning
analysis that the error  we quote is probably slightly
underestimated at the lower $T$ values.   So we find this result
reassuring, consistent with the serious critical slowing down
that we are observing and with critical  fluctuations.

\begin{figure}
  \epsffile[22 206 565 690]{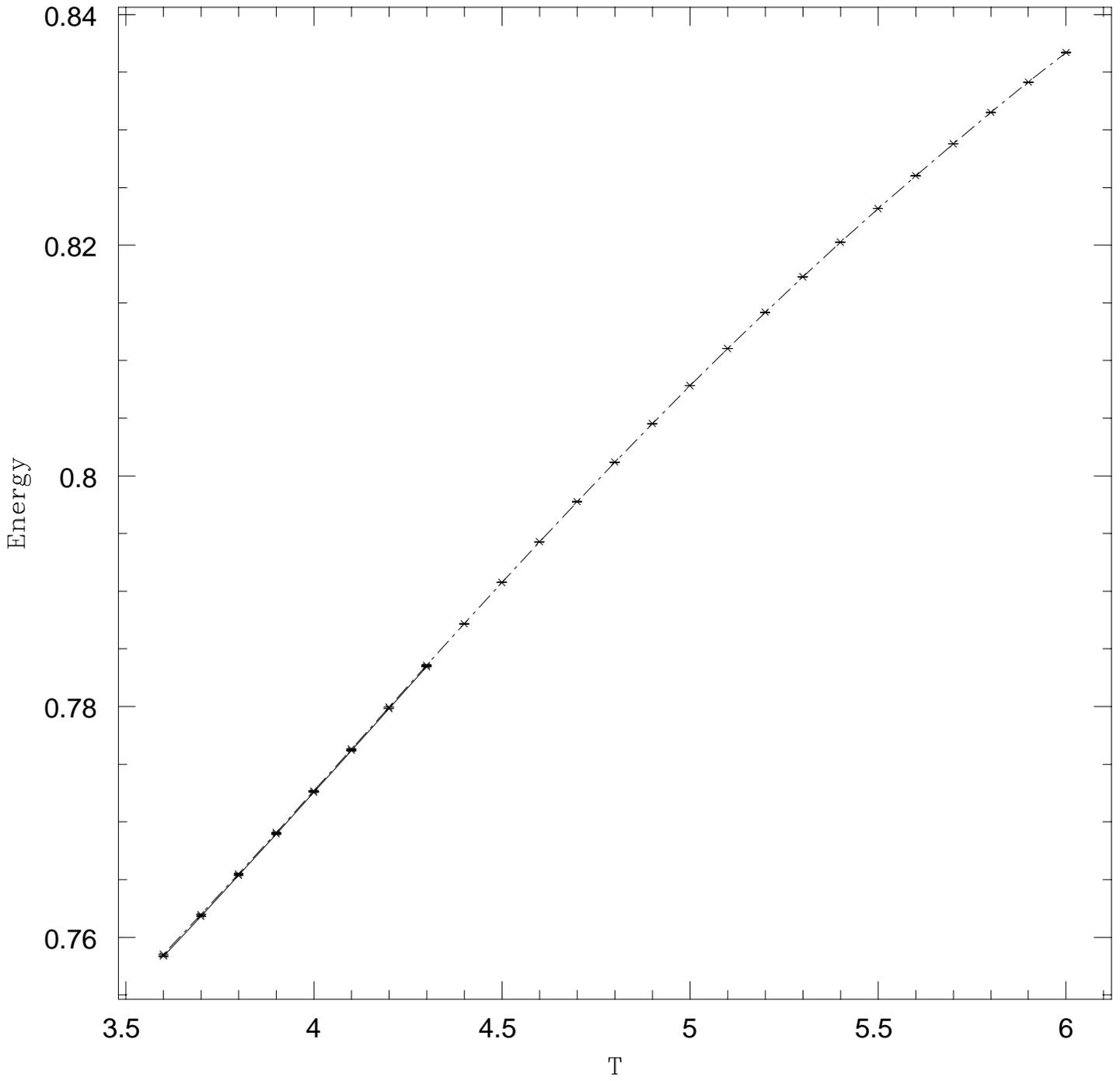}
  \caption[a]{\protect\label{F_2ENE}
  As in fig.~(\ref{F_2CHI}), but the two internal energies.}
\end{figure}

\begin{figure}
  \epsffile[22 206 565 690]{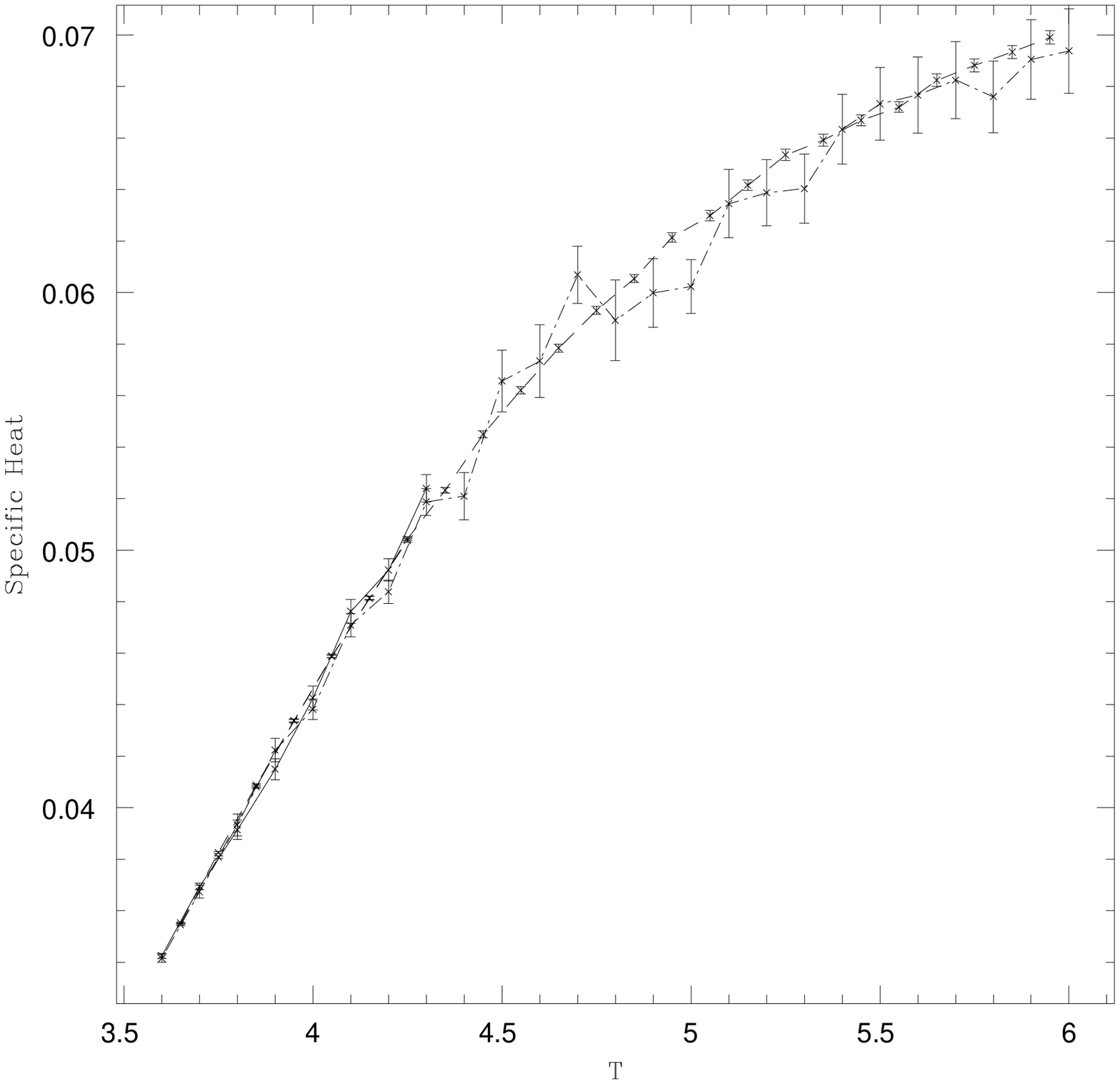}
  \caption[a]{\protect\label{F_2CV}
  As in fig.~(\ref{F_2CHI}), but the specific heats.  The point with a
  large statistical noise are from the energy fluctuations, while the
  ones with a smaller noise are from $T$ derivatives of the internal energy.}
\end{figure}

The internal energies of the two systems are completely compatible
(fig. (\ref{F_2ENE})), as it is the specific heat (which we measure
both from equilibrium fluctuations and from the $T$ derivative of the
internal energy, fig.~(\ref{F_2CV})).  We feel confident that on the
$64\times 64\times 128$ lattice results do not vary much with the
sample, and in the following we will discuss results averaged over the
two realizations of the quenched disorder.

We have estimated statistical errors by a binning analysis. We have
systematically blocked the data in coarser and coarser sub-samples, to
check statistical independence of the configuration groups eventually
used for the final error analysis.  Always but for the two lower $T$
values ($3.6$ and $3.7$) we have reached a very reliable estimate of the
true statistical error. In the two last cases the error seems
stabilizing under binning, but the evidence is less compelling, and we
would allow for a possible small underestimation of the statistical
error (of less, say, than $50 \%$).

For $T$ going from $6.0$ down to $4.4$ we present errors based on $9$
blocks of order of $50,000$ configurations (the actual measurements were
taken just once in $200$ sweeps). From $4.3$ down to $3.8$ we have $9$
blocks of order $400,000$ configurations each. At $T=3.7$ we have used
$5$ blocks of $3\times 10^6$ configurations, and at $T=3.6$ $6$ groups
of $6.5 \times 10^6$ configurations.

\begin{figure}
  \epsffile[22 206 565 690]{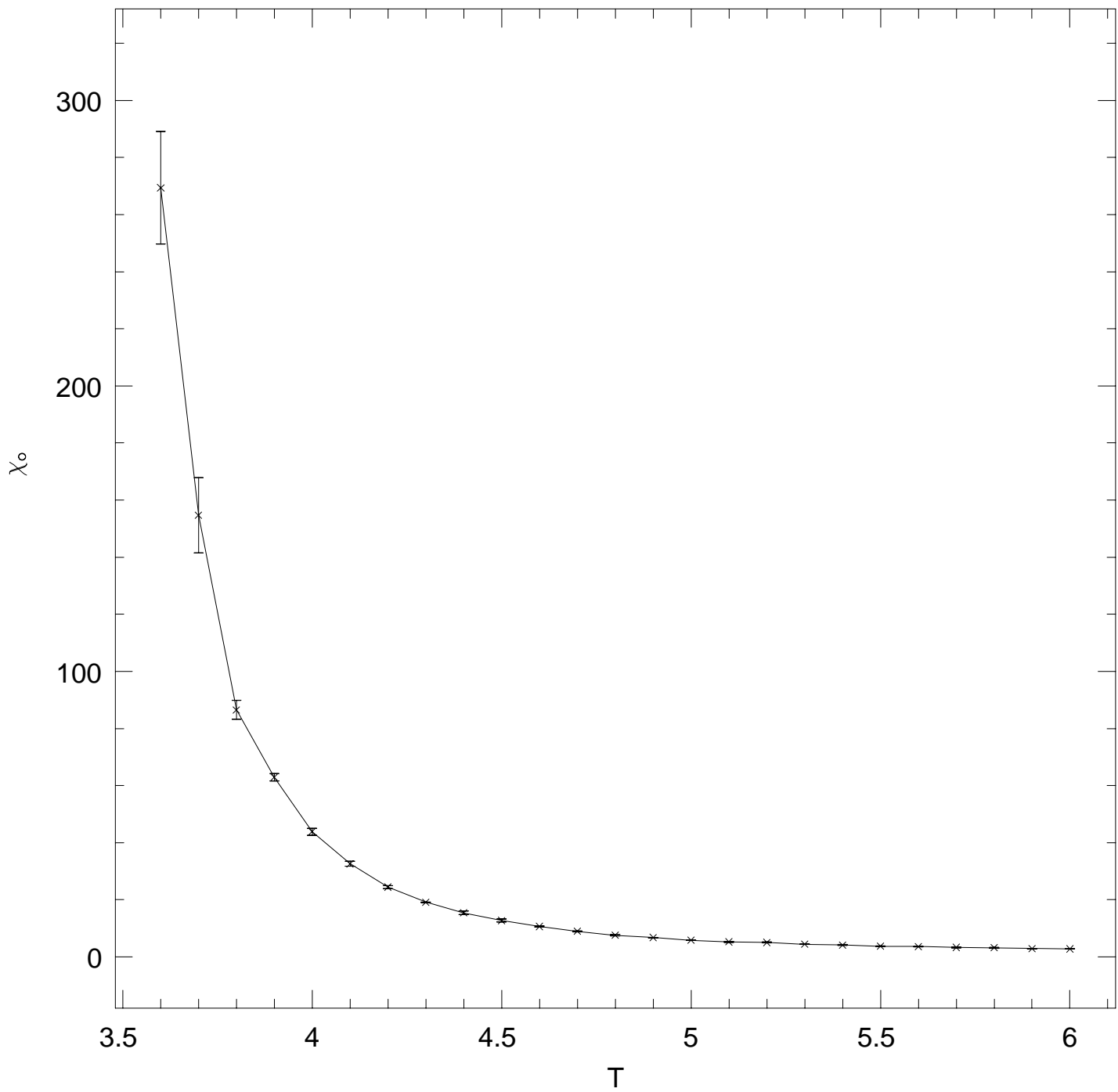}
  \caption[a]{\protect\label{F_CHIO}
  The overlap susceptibility, averaged over the two different
samples.}
\end{figure}

In fig.~(\ref{F_CHIO}) we plot the final overlap susceptibility,
averaged over the two coupling realizations, as a function of the
temperature $T$.

Our main goal has been trying to establish (or disprove) the existence
of a finite $T$ phase transition for the $3d$ spin glass model under
study. Since correlation times diverge very fast when approaching the
low temperature region (or $T_c^+$, if it exists), we are not in an
easy situation. On a large lattice we have to look at data far away in
the warm phase (the one we can check and trust have thermalized), and
try to decide which kind of critical behavior they have.

\begin{figure}
  \epsffile[22 206 565 690]{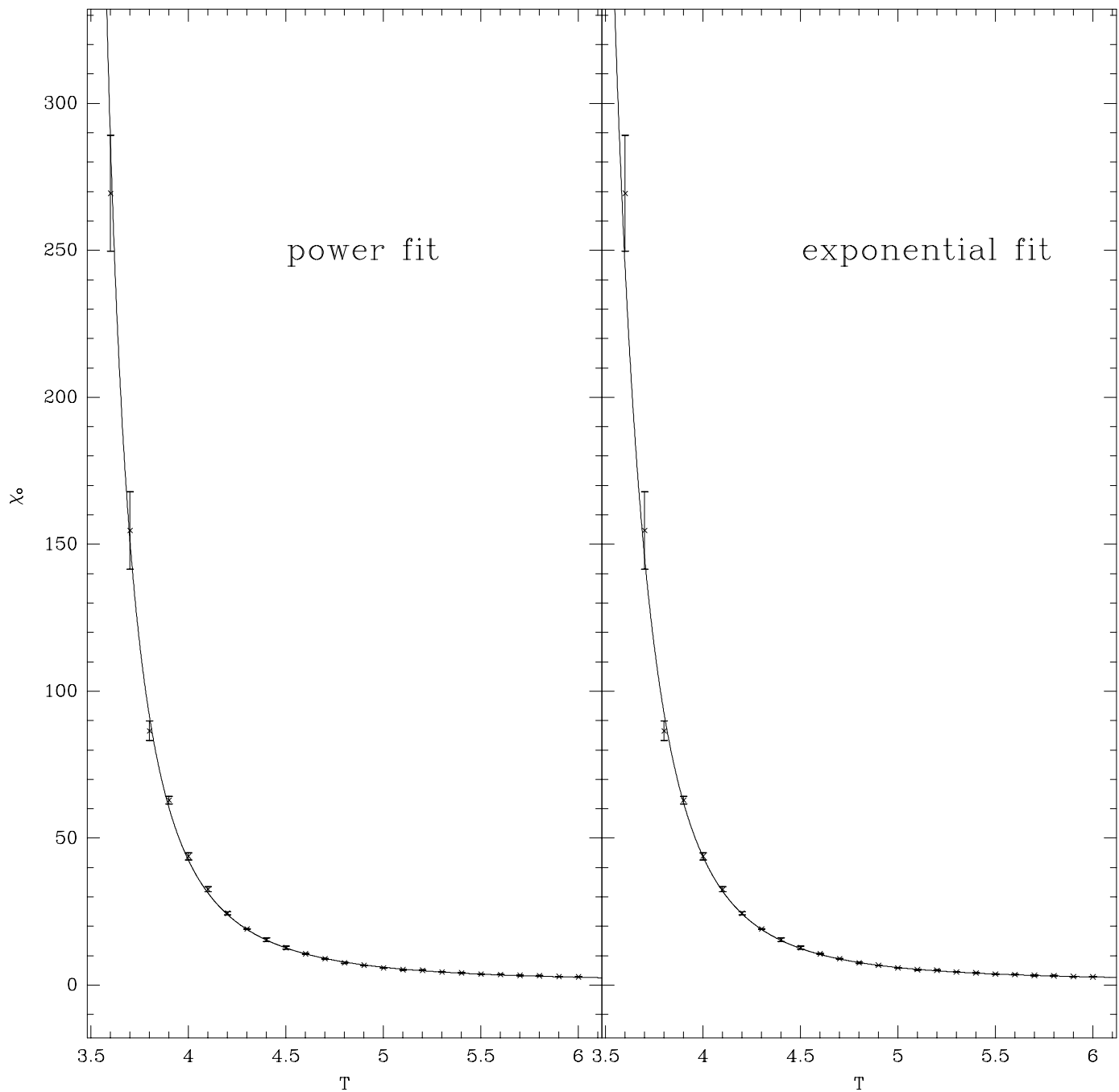}
\caption[a]{\protect\label{F_CHIFIT}
The overlap susceptibility,  same than in fig.~(\ref{F_CHIO}).  Here
the continuous lines are the results of the best fits. In
fig.~(\ref{F_CHIFIT}a), on the left, the result of the power fit to
the form (\ref{E_CHIPOW}), and in fig.~(\ref{F_CHIFIT}b), on the
right, the result of the power fit to the form (\ref{E_CHIEXP}). }
\end{figure}

At first we have tried fitting $\chi_o$ with a power divergence at the
critical temperature $T_c$, i.e.,

\be
 \protect\label{E_CHIPOW}
 \chi_o \simeq 1 + \frac{A^{\chi}_p}{(T-T_c)^\gamma}\ ,
\ee

\noindent where the subscript $p$ stands for {\em power} fit. We
show in
fig.~(\ref{F_CHIFIT}a) our best fit, obtained by using all the data
points shown in the figure. The results are

\be
  \protect\label{E_CHIPARPOW}
  A^{\chi}_p = 19.3 \pm 1.1,\  T_c = 3.27 \pm .02,\  \gamma = 2.43
\pm .05\ .
\ee

\noindent We do not attach much significance to the statistical
errors quoted here. They are reasonable estimates of a standard
fitting routine, but not the result of a detailed study of a very
complex $3$-parameter fit. We will see in a moment that the main issue
here is not the statistical error, but the systematic error, which is,
as far as we can judge from the present data, infinite (see later).

Obviously one would like to select a $T$ region that would allow
exposing a good scaling behavior (and to be obliged from the fit to
discard a high $T$ region where scaling corrections are important and a
region close to $T_c$ where finite size effects become sizeable).  This
would amount, in some sense, to find at least the size of the first
corrections to scaling. In the present case we have to compromise on the
quality of the results in (\ref{E_CHIPARPOW}), which is, still,
reasonably good. We have checked that by fitting only points close to
$T_c$ we get results that are not so different from the ones given in
(\ref{E_CHIPARPOW}). For example if we fit from $T=5.0$ down to $T=3.6$
we obtain $\gamma = 2.67 \pm .06$, and $T_c=3.20$.

Let us repeat that here the problem
will turn to be mainly the systematic error.

The second functional behavior we have tried assumes no critical
point, but an essential singularity at $T=0$. We have first tried
the form

\be
  \chi_o \simeq A^{\chi}_e ( e^{(\frac{B^{\chi}_e}{T})^P} - 1 )
  + C^{\chi}_e \ ,
\ee

\noindent where the subscript $e$ stands for {\em exponential} fit.
The power $P$ turned out to be very close to $4$ (also for the
exponential fit to the correlation length $\xi$, see later). We have
tried fits with different fixed power $P$, and for the fit to $\chi_o$
(the fit to $\xio$ has a larger indetermination, see later) we find
that a power of $3$ or $5$ gives clear worse results than a power $4$.
So we have eventually used the $3$-parameter fit to the form

\be
  \protect\label{E_CHIEXP}
  \chi_o \simeq A^{\chi}_e ( e^{(\frac{B^{\chi}_e}{T})^4} - 1 )
  + C^{\chi}_e \ ,
\ee

\noindent which gives results

\be
  \protect\label{E_CHIPAREXP}
  A^{\chi}_e = 1.67 \pm .05,\  B^{\chi}_e = 5.38 \pm .01\ ,
  C^{\chi}_e = 1.28 \pm .05,\
\ee

\noindent The best fit is very good, and we show it in
fig.~(\ref{F_CHIFIT}b), on the right. The $\chi^2$ is much better than
for the power fit ($12$ versus $29$ with some slightly arbitrary
normalization).

\begin{figure}
  \epsffile[22 206 565 690]{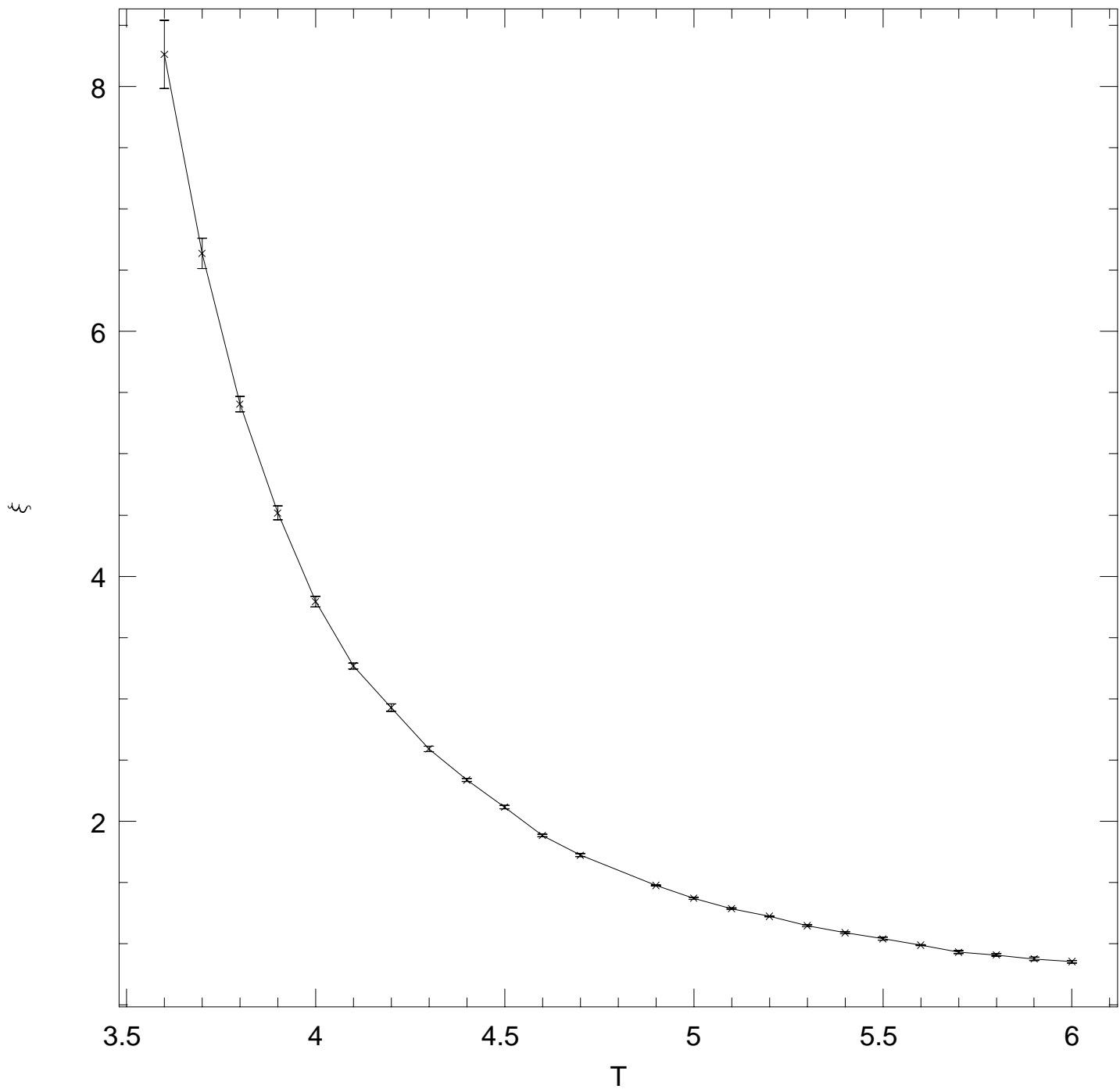}
\caption[a]{\protect\label{F_CSI1} The correlation length $\xio$,
averaged over the two different samples. The continuous line is
here
only to join neighboring points.}
\end{figure}

The divergence of the correlation length as a function of $(T-T_c)$
gives, if a phase transition exists, the exponent $\nu$. We have
repeated here the analysis we have  discussed for $\chi_o$. In
fig.~(\ref{F_CSI1}) we give $\xio$ (we have defined before) as a
function of $T$. $\xiz$ is always compatible with $\xio$, but has a
larger statistical error.

Our estimator for $\xio$ is defined by taking the weighted average
of the effective mass estimator at distance $d$

\be
  \tilde{m}(d) \equiv \log(\frac {C^{(1)}(d)} {C^{(1)}(d+1)})\ ,
\ee

\noindent for $d$ going roughly from $\xi$ to $2 \xi$. In this way
we are making systematic effects (coming from small distance
contributions) and statistical error small. A typical fitting window
is $d$ from $2$ to $3$ at large $T$ down for example to $8$ to $15$ at
$T=3.7$. We have estimated errors by using a standard binning plus
jack-knife procedure. Our conclusions about the statistical
significance of the sample coincide with the ones we have drawn for
$\chi_o$.

Also in this case we have tried a power fit and an exponential fit.
For the power fit we used the form

\be
 \protect\label{E_CSIPOW}
 \xio \simeq \frac{A^{\xi}_e}{(T-T_c)^\nu}\ ,
\ee

\noindent with the result

\be
  \protect\label{E_CSIPARPOW}
  A^{\xi}_e = 2.73 \pm .11,\  T_c = 3.24 \pm .03,\  \nu = 1.20 \pm
.04\ .
\ee

\begin{figure}
  \epsffile[22 206 565 690]{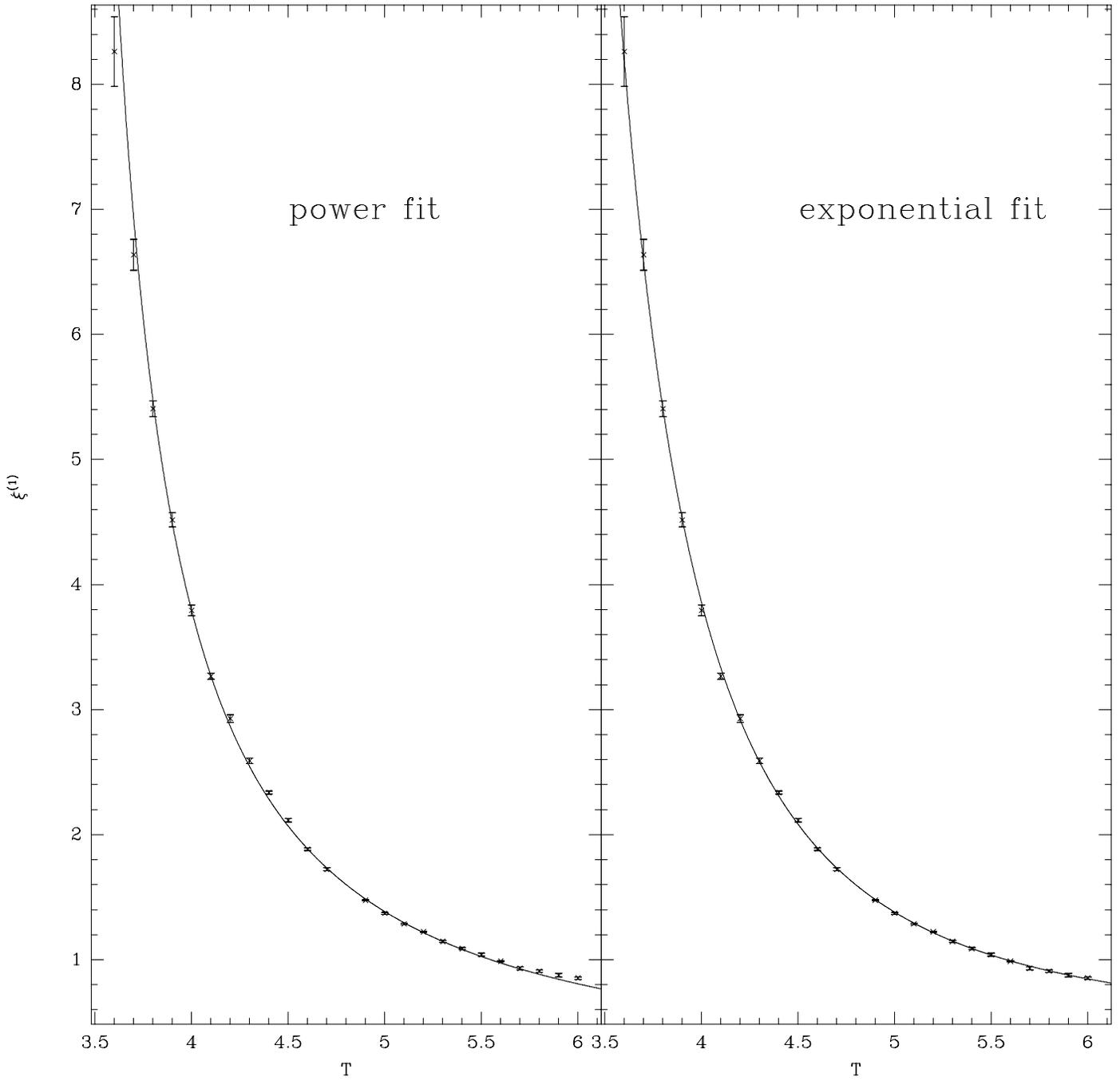}
\caption[a]{\protect\label{F_CSIFIT} The correlation length, same
than in fig.~(\ref{F_CSI1}).  Here the continuous lines are the results
of the best fits. In fig.~(\ref{F_CSIFIT}a), on the left, the result of
the power fit to the form (\ref{E_CSIPOW}), and in
fig.~(\ref{F_CSIFIT}b), on the right, the result of the power fit to the
form (\ref{E_CSIEXP}). }
\end{figure}

Even if the results are very reasonable, the fit is not good (as
shown in fig.~(\ref{F_CSIFIT}a), on the left). The $\chi^2$ is
very high ($\simeq 120$), and the points close to $T_c$ are the one
that do not fit (very dangerous {\em caveat}!). Still, if we take
these
data seriously, we have to notice that $T_c$ is the same we
estimated
by using $\chi_o$, and that by means of the scaling relation

\be
  \gamma = \nu ( 2 - \eta)
\ee

\noindent we get $\eta \simeq 0$.

The exponential fit has the form

\be
  \protect\label{E_CSIEXP}
  \xi \simeq A^{\xi}_e ( e^{(\frac{B^{\xi}_e}{T})^4} - 1 )
  + C^{\xi}_e \ ,
\ee

\noindent and gives

\be
  \protect\label{E_CSIPAREXP}
  A^{\xi}_e = 1.41 \pm .05,\  B^{\xi}_e = 4.21 \pm .02\ ,
  C^{\xi}_e =  .46 \pm .01,\ .
\ee

\noindent Such best fit is very good, and we show it in
fig.~(\ref{F_CSIFIT}b). The $\chi^2$ is $4$ times smaller than for the
power fit. This fit is by far a better fit than the fit to a power law
behavior.

For $\xi$ the evidence for the power in the exponential
being $4$ is less compelling than for $\chi_o$. Here fit with power
$2$, $3$ or $5$ are acceptable, also if the $\chi^2$ is minimum at
power $4$ (or $5$, which gives a very similar fit. For power $3$ a
small decrease in quality is already apparent).

\begin{figure}
  \epsffile[120 206 565 440]{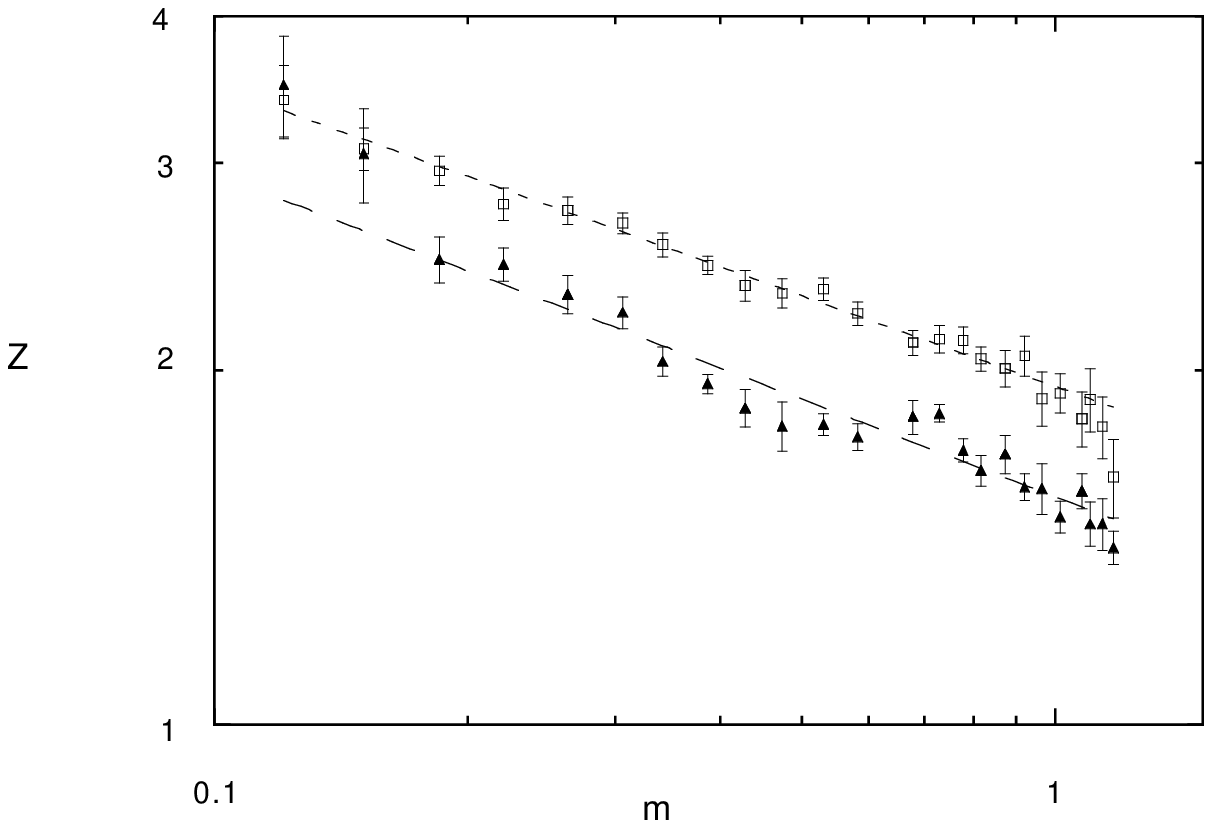}
\caption[a]{\protect\label{F_DUEZETA} The quantities $Z_0$ (lower curve)
and $Z_1$ (upper curve) as function of $m$ in a logarithmic scale.}
\end{figure}

In fig.~(\ref{F_DUEZETA}) we show the data for

\be
  Z_0 \equiv \chi_o m^2\ ,Z_1 \equiv g_1 m\ ,
\ee

\noindent from the data we have already shown for
$m$, the inverse correlation length.  We expect both quantities to
diverge as $m^{\eta}$ in the small $m$ limit. Both quantities can are
well fitted with a power law with $\eta \sim -.25$.

An independent way to measure $\eta$ is to study directly the data for
the correlation function $G^{(1)}$. At large distances the data can be
fitted as

\be
  \frac{{\cal Z}(\beta)}{r} e^{-mr}\ .
\ee

\noindent ${\cal Z(\beta)}$ seems to diverge close the  critical temperature,
with a very small power $\sim m^{-.1}$, making this estimate of
$\eta$ quite different from the previous one.  The discrepancy
among the two estimates of $\eta$ is likely to be related to the small
asymptotic value of $\eta$.

As a check we have analyzed the data for the correlation function

\be
  C(s) \sim \frac{r G^{(1)}(r)}{{\cal Z}(\beta)}\ ,
\ee

\noindent in the scaling region as function of
$s \equiv \frac{r}{\xi}$. The fact that the exponentially decaying
fits to the correlation function are good implies that for $s>1$
the function $C(s)$ is well approximated by $e^{-s}$. At small
values of $s$ the function should go to zero as $s^\eta$. Alas, since
we cannot reach very small values of $s$ it is difficult to
use this method to get a precise determination of $\eta$.

\begin{figure}
  \epsffile[120 206 565 440]{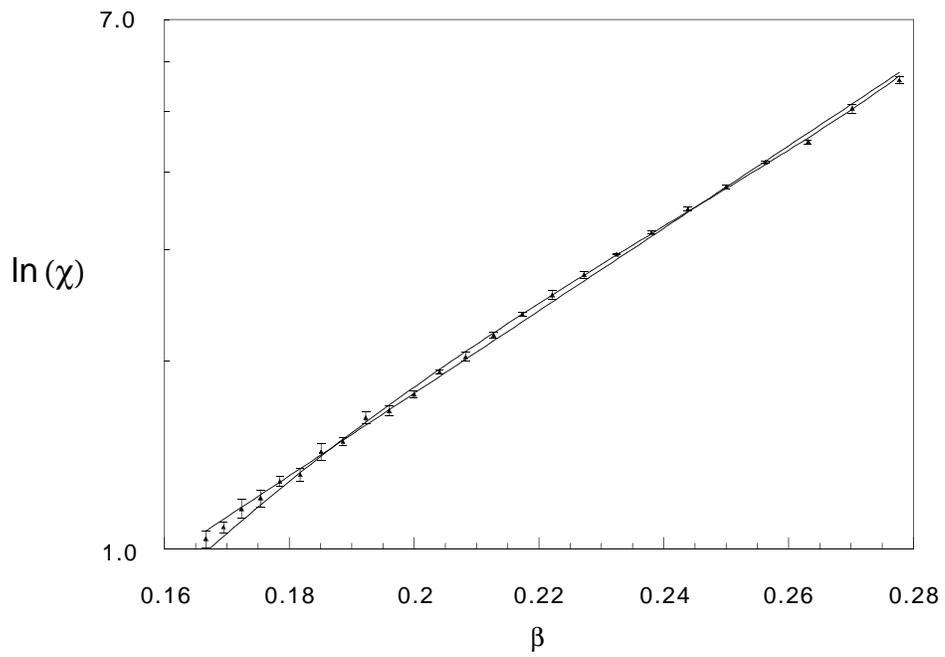}
  \caption[a]{\protect\label{F_PRIMIDUE}
  Two fits to the data for the susceptibility $\chi_o$, as a function
of $\beta$, according to  equations (\ref{E_CHIPOW},\ref{E_FITUNOB}).}
\end{figure}

\begin{figure}
  \epsffile[120 206 565 440]{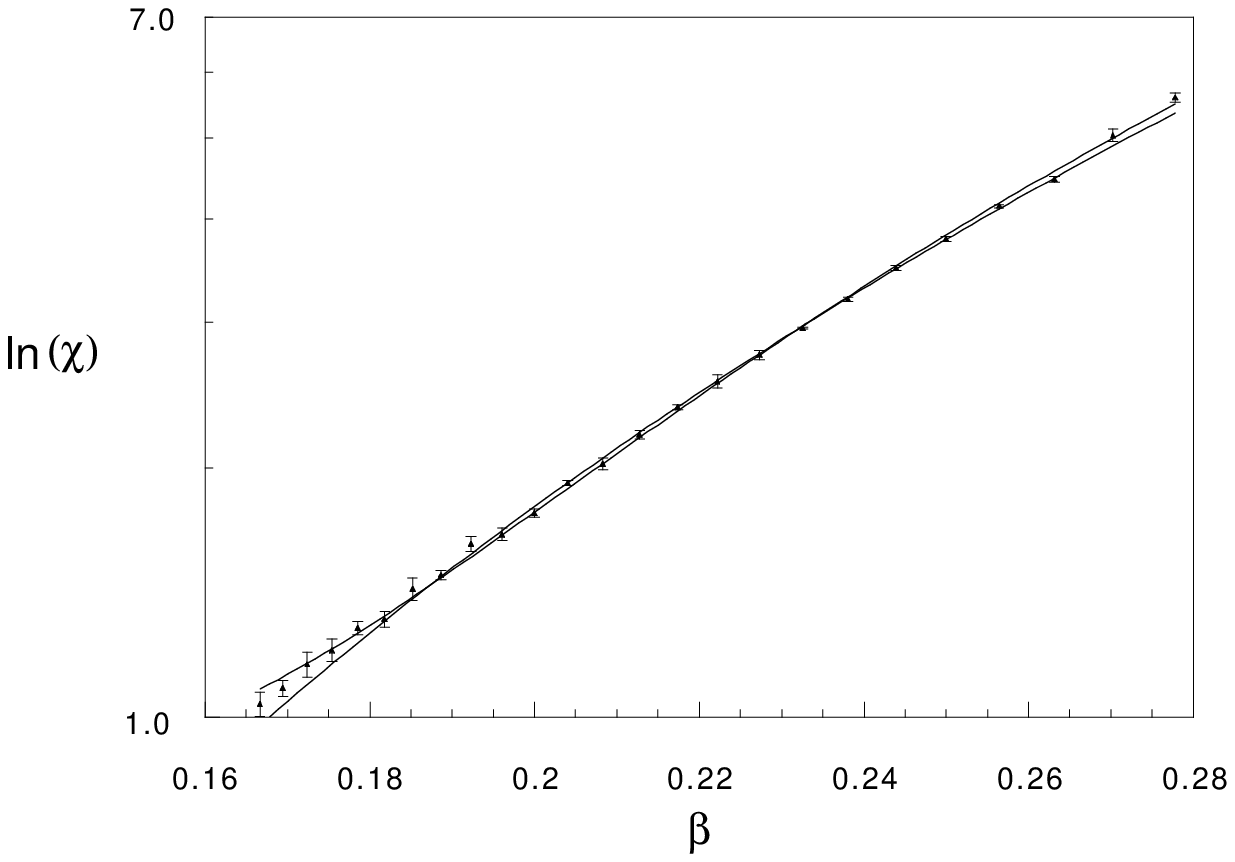}
  \caption[a]{\protect\label{F_SECONDIDUE}
  Two fits to the data for the susceptibility $\chi_o$, as a function
of $\beta$, according to equation (\ref{E_FITDUE}).}
\end{figure}

Let us insist on the difficulty in reaching a definite conclusion
about the critical regime by presenting some more fits (figures
(\ref{F_PRIMIDUE}) and (\ref{F_SECONDIDUE}). Here we are
analyzing the overlap susceptibility $\chi_o $ as function of $\beta$.
In fig.~(\ref{F_PRIMIDUE}) we show the best fit to the form
(\ref{E_CHIPOW}) with the parameters given in (\ref{E_CHIPARPOW})
(with a transition at a critical temperature), and we superimpose
a second fit of the form

\be
  \log (\chi_o)  \ = \  A \ e^{(B \beta)}\ ,
  \protect\label{E_FITUNOB}
\ee

\noindent with $ A=.085$ and $B=15.16$. Again, although
the two functional forms imply a very different critical behavior, in
the region we have studied they are indistinguishable.

We can try more.  A similar phenomenon is displayed in fig.
(\ref{F_SECONDIDUE}). Here we show dependencies that imply a
transition at zero temperature:

\bea
  \log(\chi_o) & = & A \beta^\omega \\
  \log(\chi_o) & = & A+ B \beta + C \beta^2\ .
  \protect\label{E_FITDUE}
\eea

\noindent In the first best fit we
find $A=383$ and $\omega=3.33$, while in the second best fit we
get $A=5.9$, $B=-69.8$ and $C=246$. $\omega$ turns out to be not so far from
$4$, as we already remarked.

The four fits all give reasonable results. It is impossible to use the
data to reject one of them. Of course we could choose the one with
smallest $\chi^2$, but this procedure may give an incorrect answer
since we have neglected sub-asymptotic terms, inducing an systematic
error which are out of control.

{}From these data, we tend to conclude we have a hint for the absence
of a phase transition in the $3d$ spin glass. If on the contrary
such a phase transition is present, than we have given a reasonably
precise estimate of the critical exponents.

\section{Conclusions\protect\label{S_END}}

We believe we have pointed out an open problem that in recent
papers was quoted as solved. Nowadays it is usually said that the
existence of a phase  transition is established. For example ref.
\cite{RIEGER} about aging phenomena (see \cite{AGING} for more
aging papers) claims that it is common lore that $3d$ spin
glasses undergo a finite $T_c$  phase transition.  It does
not seem to us that the existence of a phase transition is well
established at all.

The possibility of $3$ being the lower critical dimension is
appealing. We have in mind a scenario where the predictions of the
mean field theory describe fairly the behavior of the system down to
$d=3$, where the transition disappears. In no cases, as it is sensible
to expect, the system behaves as a normal ferromagnet. At low $T$ in
$3d$ the system is reminiscent of the mean field picture up to a
critical length which is function of $T$, and diverges at $T=0$.

As it was noted many years ago in ref. \cite{FOUR} at the lower
critical dimension we expect $\frac{1}{f}$ noise for the power
spectrum of the magnetization, that agrees with what has been observed
experimentally
\cite{EXPNOI}.

It is clear that there is an apparent critical temperature. Close
to this pseudo-$T_c$ the correlation length becomes so large that it
cannot be measured on the lattice sizes that are normally studied.
Below such temperature the system behaves as if it is in the
low temperature phase, irrespectively of the existence of the
transition (think about the $1d$ normal Ising model for low values
of $T$).

The only way to disprove the existence of a transition at finite
temperature would be to show that the data for the susceptibility and
the correlation length cannot be fitted with power law singularities
at finite temperature. On the contrary to present an evidence for a
transition at finite temperature one should show that the data can be
fitted as power law singularities and cannot be fitted with functions
having only singularities at zero temperature. Our data, as well those
from the very long simulations of Ogielski and Morgenstern
\cite{OGIMOR,OGIELS}, can be fitted in both ways. As we already said,
we do not think that we can discriminate the two admissible behaviors
from the value of the chi-square $\chi^2$, i.e., of the quality of the
fit, especially in an approach where corrections to scaling have been
neglected.  Unfortunately in absence of clear predictions about the
low temperature behavior it is difficult to exclude the possibility of
a transition at $T=0$.

\begin{figure}
  \epsffile[120 206 565 440]{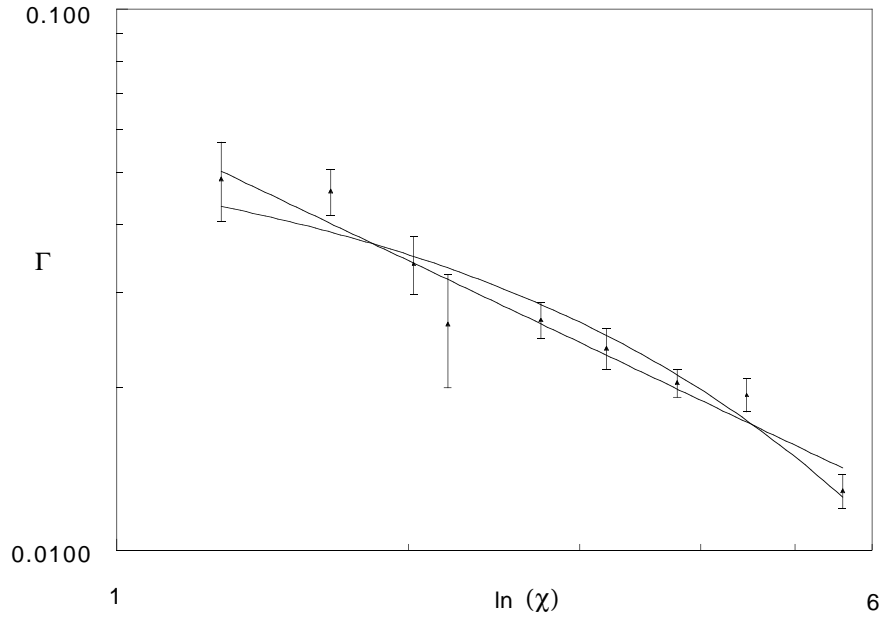}
\caption[a]{\protect\label{F_CONC} $\Gamma$ versus $\tau$ (see
the definitions in the text).}
\end{figure}

To visually discriminate among the two possibilities we
plot in figure (\ref{F_CONC}) the quantity

\be
  \Gamma \equiv \frac{d \beta}{d \log(\chi_o)}\ ,
\ee

\noindent  versus $\tau \equiv \log(\chi_o)$.
A finite $T$ transition implies that

\be
  \Gamma \simeq  e^{-A \tau}\ ,
\ee

\noindent  with $A=1/\gamma$, while a transition at $T=0$ with a
divergence of the form $\exp(\beta^\omega)$ implies

\be
  \Gamma \simeq  \frac{1}{\tau^B}\ ,
\ee

\noindent with $B=1-{1\over \omega}$. $B=1$ corresponds to an
$e^{(e^\beta)}$ behavior. Our best fits give $A=.29$ and $B=.86$.  The
data are noisy at high temperature (low $\tau$). Clearly it is
difficult to select one fit, especially since we have neglected
corrections to scaling. Data seems to prefer a straight line with a
coefficient not far from one, but we are unwilling to rely on this
kind of evidence.

\begin{figure}
  \epsffile[120 206 565 440]{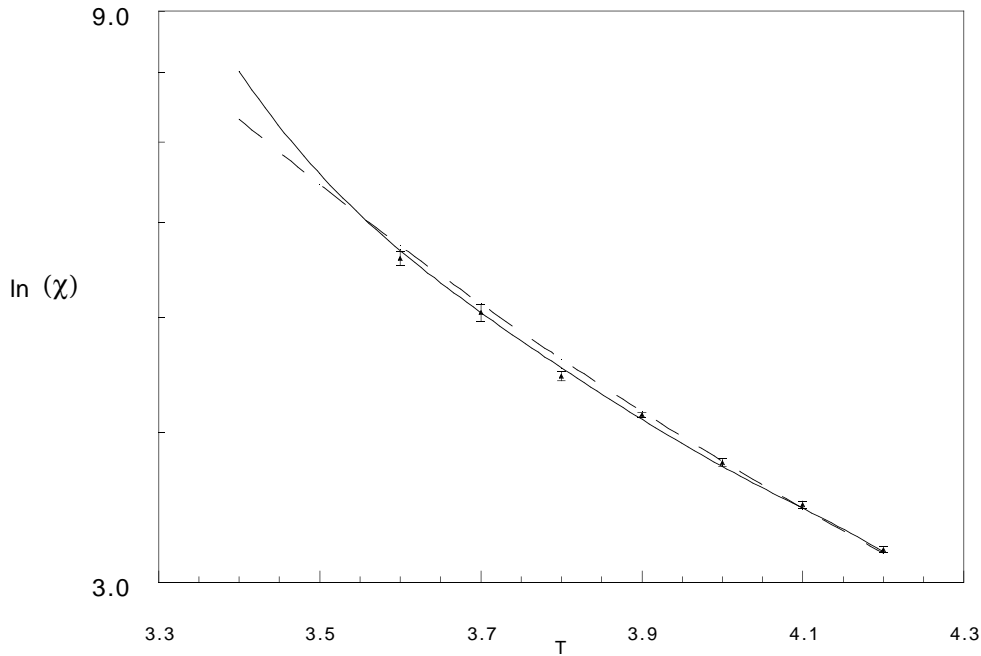}
  \caption[a]{\protect\label{F_CONF}
  $\log{\chi_o}$ versus $T$. Fit $1$ (continuous line)
  is to a power law singularity, as in eq. (\ref{E_CHIPOW})
  fit $2$ (dashed line)
  has a $T=0$ singularity, as in eq. (\ref{E_FITUNOB}). }
\end{figure}

What can be done with a better numerical simulation? To
get a hint we have extrapolated two typical fits at a reasonable
low $T$. We show them in fig. (\ref{F_CONF}).
We have considered a simple power singularity at $T \not=
0$, and a divergence at $T=0$ of the form $e^{Ae^B}$.
{}From our present best fits we can deduce that at, say, $T=3.4$,
we would be able to discriminate. If the data would be really following
the finite $T$ singularity scenario (first case), the strong
increase of the susceptibility could not be fitted by the double
exponential scenario, and the zero temperature transition should be
refuted.

In the opposite case, where the  hypothetical data would follow a
form of the second kind (a double exponential singularity at $T=0$)
we find that a power fit would still be a good fit, but with a
larger value of $\gamma$ and smaller value of $T_c$. This variation
of the value of the best fit parameters with the temperature
interval used for the fitting would then be taken as good evidence
for the existence of a zero temperature transition.

If the double exponential singularity behavior is correct, the
correlation length should increase of a factor about $2.5$ when going
from $T=3.6$ to $T=3.4$. That means that a reliable estimate will be
possible on $128^3$ lattice, only slightly larger of what we used
here, and not out of reach of the present technology. An increase of
the computer time of more than one order of magnitude seems
unfortunately necessary, but this is also a reasonable goal. Such a
computation seems possible in a not too far away future.

It is also possible that a careful analysis of the model at low
$T$ could allow to show the absence of a phase transition. In
this case it would be essential to identify the renormalization
group flow away from the zero temperature fixed point.

\section*{Acknowledgements}

This work was partially supported by the National Science Foundation
thr\-ough grant No. DMR-9217284.  F.R. has been supported by the EEC
fellowship B/SC1*/915198.  We thank Cristina Marchetti for many
interesting discussions. We have used intensively the NPAC DECmpp. We
thank Geoffrey Fox for his continuous support and for interesting
discussions.  We are grateful to Mark Levinson, the NPAC DECmpp
manager, for his really wonderful and crucial help and support.

\vfill

\end{document}